\pdfoutput=1
\documentclass[12pt]{article}
\usepackage{amsmath,amssymb}
\usepackage{dcolumn}
\usepackage{graphicx,color}
\usepackage{bbold}
\usepackage{bm}
\usepackage[colorlinks]{hyperref}
\usepackage{subfigure}
\usepackage{fmdg}

\usepackage{bigstrut}

\usepackage{hyperref}
 \hypersetup{
 pdfauthor={Michael A. Schmidt and Alexei Yu. Smirnov},
 pdftitle={Neutrino Masses and a Fourth Generation of Fermions}
 }

\usepackage[sort&compress,numbers,colon,merge]{natbib}
\newcommand{\be}{\begin{equation}}
\newcommand{\ee}{\end{equation}}

\newcommand{\bea}{\begin{eqnarray}}
\newcommand{\eea}{\end{eqnarray}}

\usepackage{bigstrut}
\newcommand{\Rep}[1]{\underline{\mbox{\textbf{#1}}}}
\newcommand{\MoreRep}[2]{\underline{\mbox{\textbf{#1}}} _{\mbox{\textbf{#2}}}}

\usepackage{amsmath,amssymb}
\usepackage{amsfonts}
\usepackage{slashed}
\usepackage{bbold}
\usepackage{bm}

\newcommand{\ev}[1]{\ensuremath{\left\langle #1 %
                     \right\rangle}} 

\newcommand{\U}[1]{\ensuremath{\mathrm{U}(#1)}}

\newcommand{\I}{\ensuremath{\mathrm{i}}}
\newcommand{\diag}{\ensuremath{\mathrm{diag}}}
\newcommand{\TeV}{\ensuremath{\mathrm{TeV}}}
\newcommand{\GeV}{\ensuremath{\mathrm{GeV}}}
\newcommand{\MeV}{\ensuremath{\mathrm{MeV}}}

\newcommand{\eV}{\ensuremath{\mathrm{eV}}}
\newcommand{\hc}{\ensuremath{\mathrm{h.c.}}}

\newcommand{\MSbar}{\ensuremath{\overline{\mathrm{MS}}}}

\newcommand{\Eqref}[1]{Eq.~\eqref{#1}}
\newcommand{\Figref}[1]{Fig.~\ref{#1}}
\newcommand{\Tabref}[1]{Tab.~\ref{#1}}
\newcommand{\Secref}[1]{Sec.~\ref{#1}}
\newcommand{\Appref}[1]{App.~\ref{#1}}

\newcommand{\dd}{\ensuremath{\mathrm{d}}}

\newcommand{\SG}[2]{\ensuremath{\mathrm{SG}(#1,#2)}}
\newcommand{\vEWnu}{\ensuremath{v_\mathrm{EW}^\nu}}
\newcommand{\vEWe}{\ensuremath{v_\mathrm{EW}^e}}

\begin{document}
\topmargin 0pt
\oddsidemargin=-0.4truecm
\evensidemargin=-0.4truecm
\renewcommand{\thefootnote}{\fnsymbol{footnote}}


\begin{titlepage}

\ \vspace*{-15mm}
\begin{flushright}
IPPP/11/51\\
DCPT/11/102
\end{flushright}
\vspace*{5mm}

\begin{center}
{\huge\sffamily\bfseries 
Neutrino Masses and a Fourth Generation of Fermions
}
\\[10mm]
{\large
Michael A.~Schmidt\footnote{\texttt{michael.schmidt@unimelb.edu.au}}$^{(a)(b)}$ and
Alexei Yu.~Smirnov\footnote{\texttt{smirnov@ictp.it}}$^{(c)}$}
\\[5mm]
{\small\textit{$^{(a)}$
Institute for Particle Physics Phenomenology, Durham University,
Durham DH1 3LE, UK
}}
\\
{\small\textit{$^{(b)}$
ARC Centre of Excellence for Particle Physics at the Terascale,
School of Physics, The University of Melbourne, Victoria 3010, Australia
}}
\\
{\small\textit{$^{(c)}$
International Centre for Theoretical Physics, Strada Costiera 11, 34014 Trieste, Italy
}}
\end{center}
\vspace*{1.0cm}
\date{\today}

\begin{abstract}
\noindent We study neutrino mass generation in models with
four chiral families of leptons and quarks and four right handed
neutrinos. Generically, in these models there are three different
contributions to the light neutrino masses:
the usual see-saw contribution, the tree-level contribution
due to mixing of light  neutrinos with neutrino
of the fourth generation, and the two loop contribution
due to the Majorana mass term of the fourth neutrino. 
We study properties of these contributions and
their experimental  bounds. The regions of the parameters
(mixings of the fourth neutrino, masses of RH neutrino components, etc.)
have been identified  where various  contributions dominate.
New possibilities of a realisation of the flavour symmetries in the
four family context are explored. In particular, we
consider applications of the smallest groups, e.g. SG(20,3), with
irreducible representation \Rep{4}.
\end{abstract}


\end{titlepage}

\newpage
\setcounter{footnote}{0}

\section{Introduction}

There are various arguments in favour of the existence of  a fourth Standard Model (SM)
generation of fermions. 

- A fourth generation can alleviate the tension between the lower bound on the
Higgs mass from LEP II and the fit of the electroweak precision data,
which predicts a light Higgs particle~\cite{He:2001tp,*Alwall:2006bx,*Kribs:2007nz,*Novikov:2009kc,*Hashimoto:2010at,*Erler:2010sk,*Chanowitz:2010bm,*Baak:2011ze}. 
Indeed, the mass splittings between the fourth generation fermions can lead to a
negative contribution to the $S$ parameter, which allows for a heavier Higgs.
The flavour sector with four families has been thoroughly
analysed in~\cite{Buras:2010pi,*Hou:2010mm,Buras:2010nd,Eberhardt:2010bm,Alok:2010zj,*Hou:2011fw}. 

-  The enlarged CKM matrix contains additional CP phases and naturally leads
to more CP violation which can explain the deviations from the
predicted SM values in some measurements in B
physics~\cite{Hou:2005yb,*Hou:2006mx,*Soni:2010xh,*Soni:2008bc,Buras:2010pi,*Hou:2010mm}. 

- A fourth generation has been suggested as an explanation 
of the anomalous like-sign di-muon charge asymmetry~\cite{Choudhury:2010ya}.

-  A fourth generation makes viable electroweak
baryogenesis, which is not possible in the SM.
The introduction of a further
generation leads to additional CP violating phases in the quark mixing
matrix (CKM matrix), which are not constrained by experiment yet, {\it e.g.}~see \cite{Hisano:2011nx} for an analysis of the neutron electric dipole moment, and can
lead to a large enough CP violation~\cite{Hou:2008xd}. 
In addition, it has been shown,
that a strong first order phase transition is possible within the
SUSY version of a model with four SM generations (SM4)~\cite{Ham:2004xh,*Fok:2008yg} as
well as in a strongly coupled version with dynamical breaking of the
electroweak symmetry~\cite{Kikukawa:2009mu}. 

- Being similar to top quark condensate 
models~\cite{Nambu:1988mr,*Miransky:1988xi,*Miransky:1989ds,*Marciano:1989mj,*Marciano:1989xd,*Bardeen:1989ds}, 
dynamical electroweak symmetry breaking is possible in the context of four generations
~\cite{Holdom:1986rn,*Hill:1990ge,*Carpenter:1989ij,*Hung:2009hy,*Delepine:2010vw,*Hashimoto:2010fp,*Hung:2010xh,*Fukano:2011fp,BarShalom:2011zj}.

- A fourth generation neutrino can contribute to the dark matter
density of the Universe if an additional $B-4L_4$ symmetry is introduced protecting the fourth generation neutrino from decaying and 
 it couples to the three light generations via the new $Z^\prime$ to quarks~\cite{Lee:2011jk}.

-  Under the assumption of minimal flavour violation, a fourth generation suppresses proton decay and enforces the R-parity in the context of the MSSM due to the mismatch of numbers of flavours and colours~\cite{Smith:2011rp}.

Significant interest to  the fourth generation is also revived due to operation of  the
Large Hadron Collider (LHC).   
The LHC  can provide a critical test of existence of the fourth generation: 
either discover or exclude it. (See~\cite{Holdom:2009rf} for a recent review
and~\cite{Frampton:1999xi} for an earlier review.)
Indeed, the LHC can test the region of
fourth generation quark masses, $300-800$ 
GeV,~\cite{Arik:1996qd,*:1999fr,*Holdom:2006mr,*Ciftci:2008tc,*Burdman:2008qh,*Holdom:2011uv},
which covers the complete parameter space determined  by the partial wave
unitarity upper limit of $550\,\GeV$ for a quark
doublet~\cite{Chanowitz:1978uj,*Chanowitz:1978mv} and the limit  
obtained in models of a strongly coupled fourth generation. The CMS Collaboration 
put a lower bound on the mass of fourth generation
up-type quark $t^\prime$ of  $m_{t^\prime}\gtrsim450\,\GeV$~\cite{PAS-EXO:2011-005,*PAS-EXO:2011051} and exclude fourth generation down-type quark $b^\prime$ in the mass region $255\,\GeV<m_{b^\prime}<361\,\GeV$ at $95\%$ C.L.~\cite{Chatrchyan:2011em}.  
Existence of the fourth generation chiral leptons without fourth generation quarks 
looks rather unnatural and in fact this will require further complication of model 
to cancel the anomalies. 

The parameter space of the fourth generation can also be probed by looking for 
the Higgs signals~\cite{Gunion:2011ww} (this  has also been studied in the
MSSM~\cite{Cotta:2011bu}). Currently, the Higgs boson with mass $m_H$ in SM4 with
one Higgs doublet is excluded in the region $120\,\GeV<m_H<600\,\GeV$ at 95\%
C.L.~by CMS~\cite{PAS-HIG:2011-011} and $140\,\GeV<m_H<185\,\GeV$ by
ATLAS~\cite{Aad:2011qi}. However, these bounds only apply in the minimal SM4
model with one Higgs doublet. They are weakened if
(i) the Higgs production via gluon fusion is modified, e.g. by a colour
octet~\cite{He:2011ti} or if the light Higgs in a two Higgs doublet model does not couple to the fourth generation~\cite{Borah:2011ve}, (ii) the search channels $h\rightarrow WW^*, ZZ^*$ are
modified, e.g. in a two Higgs doublet model~\cite{BarShalom:2011zj}, or (iii)
the Higgs decays dominantly invisibly, e.g. into a light scalar, which can provide a dark matter candidate~\cite{He:2011de} or into fourth generation neutrinos
for light Higgs with $m_H<170\GeV$~\cite{Belotsky:2002ym,*Cetin:2011fp}. Recently, the complete
electroweak two-loop corrections to Higgs production via gluon fusion have been
calculated and discussed in the framework of a fourth
generation~\cite{Passarino:2011kv}. In the SM with four generations and one Higgs doublet, the Higgs bounds can be translated in a bound on the fourth generation fermion masses via the triviality and stability bounds~\cite{Wingerter:2011dk}.

Collider signals of the 4th generation have been
studied which include  signals of  fourth generation quarks~\cite{Arik:1996qd,*:1999fr,*Holdom:2006mr,*Ciftci:2008tc,*Burdman:2008qh,Aaltonen:2009nr,*CDFHeavyTop:2008,*Holdom:2010fr,Abazov:2011vy,Gunion:2011ww}, 
leptons~\cite{Carpenter:2010bs,*Carpenter:2010sm,Carpenter:2010dt, Rajaraman:2010wk}, sleptons~\cite{Cakir:2011dx},  signals of a
strongly coupled generation~\cite{Holdom:2010za,*BarShalom:2010bh} and a $Z^\prime$~\cite{Ari:2011is}. 
If the mixing of the fourth generation with the
three SM generations is tiny, the particles of the  fourth generation become 
long-lived~\cite{Murayama:2010xb,*Keung:2011zc}. The
fourth generation quarks might even form long-lived bound states which
can be  produced at the LHC. The binding energies and sizes of those bound
states have been calculated in~\cite{Ishiwata:2011ny}. 

Note that SM4 is  constrained by the large Yukawa couplings
running into Landau poles due to a quick renormalisation group (RG) evolution. The current experimental bounds require a cutoff scale $\Lambda_c\lesssim(10^2 - 10^{3})\,\TeV$, unless there is a fixed point~\cite{Knochel:2011ng,*Ishiwata:2011hr}. Similar results have been obtained in the SUSY context in~\cite{Holdom:2009rf,Godbole:2009sy,*Dawson:2010jx}.

Properties of the fourth generation particles should differ from the properties of three  known generations. The bound from the invisible $Z$ decay width forbids further light generations, especially additional neutrinos.  
The existing experiments give lower bounds on masses of fourth generation leptons 
(charged lepton and neutrinos) at the level of 100 GeV and an upper bound on the
mixing parameters  $0.04 - 0.08$. 

The generation of neutrino mass in models with four fermion generations
has been explored in several publications. 
The simplest model with four SM generations and usual massless  neutrinos 
at tree level has one  right-handed (RH) 
neutrino~\cite{King:1992qr,*King:1992bk}. An explanation of neutrino
masses in terms of the usual see-saw
mechanism~\cite{Minkowski:1977sc,*Yanagida:1980,*Glashow:1979vf,*Gell-Mann:1980vs,*Mohapatra:1980ia}, however, requires at least three RH neutrinos. 
Since the fourth generation neutrino should be much heavier than the
three SM neutrinos, several authors suggested its  pseudo-Dirac 
nature~\cite{Hill:1989vn,King:1992qr,*King:1992bk,Hou:2005fp,*Hou:2010qx}.
The light neutrino masses can be generated by two loop diagrams with 
two $W$ bosons exchange in the framework of five SM generations~\cite{Aparici:2011nu}. 
There are studies of the leptonic flavour structure in SM4 with 
discrete flavour symmetries $\mathbb{Z}_4$~\cite{SilvaMarcos:2002bz} and 
$A_5$~\cite{Chen:2010ty}. Moreover, the leptonic flavour structure  have been 
explored in extra-dimensional 4 generation models~\cite{Burdman:2009ih,*Lebed:2011zg}.

The origin of neutrino masses or the flavour structure, and especially the
number of chiral SM generations are  not understood within the Standard Model. 

In this paper, we present  a comprehensive study of  
the neutrino mass generation in the SM model with 
four fermionic generations including one RH neutrino per generation. 
We will restrict ourselves to a non-SUSY model. However, our results can be directly generalised to the SUSY case. Besides the usual see-saw contribution,
we calculate and study the contributions from tree level mechanism related to mixing of the 
fourth neutrino with  the light ones,  and  from  the two $W-$boson exchange at two loop.  
We explore possible flavour symmetries in the context of SM4. We study the smallest 
discrete group with a four-dimensional representation and explore flavor structures that appear in the most economical scenarios. 

The paper is organised as follows. The  contributions to the neutrino mass matrix
from three different mechanisms are computed in \Secref{sec:NuMass}. 
In \Secref{sec:discussion}, we consider  existing
bounds on fourth generation leptons, in particular, 
from the neutrinoless double beta
decay. We find  the  regions in parameter space in which different mechanisms dominate.
We explore  possible realisations of flavour symmetries in context of SM4 
in \Secref{sec:symmetries},  and conclude in \Secref{sec:conclusions}. 
In the Appendix we present the group theoretical details of the smallest group with a representation $\Rep{4}$: $\SG{20}{3}\cong\mathbb{Z}_5\rtimes_\varphi\mathbb{Z}_4$. 

\section{Contributions to Neutrino Masses\label{sec:NuMass}}

We consider the extension of the SM by one RH singlet fermion per
generation,  $N_k$, $k=1,2,3$ for the first
three generations and $N_4$ for the fourth generation. We consider two SM Higgs doublets, one coupling to neutrinos $H_1$ and one to charged leptons $H_2$. The case with a single Higgs doublet is obtained by identifying $H=H_1$ and $H^C=H_2$. In the flavour
basis, the leptons have the following couplings
\begin{equation}
-\mathcal{L}= Y_{\alpha k} \bar\ell_\alpha H_1 N_k 
+Y_{\alpha 4} \bar\ell_\alpha H_1 N_4
+Y_{E k} \bar\ell_E H_2 N_k
+Y_{E4} \bar\ell_E H_2 N_4
+\frac12 M_k N_k^T N_k +\frac12 M_4 N_4^T N_4+\hc , 
\end{equation}
where $\ell_\alpha$, $\alpha=e,\,\mu,\,\tau$ and $\ell_E$ denote
the light and the fourth generation left-handed lepton doublets, respectively. 
These couplings lead to the following neutral
fermion mass matrix in the flavour basis
$\begin{pmatrix}\nu_\alpha&\nu_E&N_4&N_k\end{pmatrix}^T$
\begin{equation}
\label{eq:NeutralFermion}
\mathcal{M}=
\left(\begin{array}{cccc}
0 &0 &  f_L & m \\
...& 0 &  m_{E4}  & f_R^T \\
... & ... & M_4& 0 \\
... & ... & ... & M \\
\end{array}\right)\; .
\end{equation}
We take  the RH
neutrino mass matrix to be diagonal: $M=\diag(M_1,\,M_2,\,M_3)$.  The complete Dirac
mass matrix consists of the following components: 

-  the Dirac mass matrix of the light
SM neutrinos, $m_{\alpha k}=Y_{\alpha k} \vEWnu$
($\alpha=e,\,\mu,\,\tau$ and $k=1,2,3$) with $\vEWnu=\ev{H_1}$ being the vacuum
expectation value (VEV) of the Higgs $H_1$ coupling to neutrinos, 

- the Dirac mass of the fourth generation:
$m_{E4}=Y_{E4}\vEWnu$, and 

- the column $f_L$ which  gives mixing of the fourth generation
with the light ones $f_{L\alpha}=Y_{\alpha 4}\vEWnu$,  and $f_{Rk}=Y_{Ek}\vEWnu$.  

The neutrino Dirac mass can be written as
\begin{equation}
U_{L}\,\diag\left(m_{i}\right)\, U_R^\dagger\;.
\end{equation}

According to (\ref{eq:NeutralFermion}) the left-handed 
(LH) mixing matrix elements of the fourth generation with the three light
generations are  given approximately by 
\begin{equation}
\left(U_{L}\right)_{\alpha 4}\simeq \frac{f_{L\alpha}}{m_{E4}}, 
\end{equation}
which are bounded by experiments to be smaller than $0.04 - 0.08$~\cite{Lacker:2010zz,*Menzel:2011ff}. 
Similarly, the RH mixing matrix elements can be estimated as 
\begin{equation}
\left(U_{R}\right)_{k 4}\simeq \frac{f_{Rk}}{m_{E4}}\;,
\end{equation}
provided that  $m_{E4}$ dominates the Dirac mass matrix.

Decoupling of the RH neutrinos, $N_{1,2,3}$, in  \Eqref{eq:NeutralFermion}
leads to the effective
mass matrix in the basis $\begin{pmatrix}\nu_\alpha,&\nu_E,&N_4\end{pmatrix}^T$: 
\begin{equation}
\label{eq:EffNeutralFermionMassMatrix}
\mathcal{M}^\prime=
\left(\begin{array}{ccc}
- m M^{-1}m^T & -
 m M^{-1} f_R  & f_{L}\\
... & -f_R^T M^{-1} f_R & m_{E4}\\
... & ... & M_{4}\\
\end{array}\right)\;.
\end{equation}
At this level the three active  neutrinos acquire the usual see-saw contributions  
associated to the three heavy RH neutrinos. 
Notice that in the limit $f_R = 0$ further  decoupling of $\nu_E$ and $N_4$ 
in \Eqref{eq:EffNeutralFermionMassMatrix} gives zero
contribution to the light neutrino masses in spite of
the fact that $\nu_\alpha$ interacts with $N_4$.

\subsection{Tree-level Mechanism from Mixing with Fourth Generation}

Depending on value of $M_4$
there are two extreme cases for the masses of  the fourth neutrino: 
the see-saw case, $M_{4} \gg m_{E4}$
and the pseudo-Dirac case $M_{4}\ll m_{E4}$.

1). In the see-saw case  after decoupling the fourth RH neutrino 
we obtain the $4\times4$ neutral fermion mass
matrix in the basis  $(\nu_\alpha, \nu_E)$: 
\begin{equation}
\left(\begin{array}{cc}
-m M^{-1}m^T- \frac{f_L^Tf_L}{M_4} & -\frac{m_{E4}}{M_4} f_L  \\
... & -\frac{m_{E4}^2}{M_{4}}\\
\end{array}\right)
\end{equation}
where we have neglected the see-saw contributions of the first three RH neutrinos 
to the fourth row and 
column since $M_4\ll M_k$. The mixing matrix elements can be estimated as 
\begin{align}
U_{\alpha 4}&\simeq \frac{f_{L\alpha}}{m_{E4}}\simeq (U_L)_{\alpha 4}\;.
\end{align}
Further  decoupling of $\nu_4$ leads to  cancellation of the leading order contribution of
the fourth generation to the light $3 \times 3$ neutrino mass matrix. 
Non-zero contributions are generated by the  next-to-leading order see-saw
effect~\cite{Schechter:1981cv,*Grimus:2000vj,*Hettmansperger:2011bt}.
Therefore, in this limit, a fourth generation does not give a substantial tree-level 
contribution to the light neutrino mass. 


2) In the pseudo-Dirac case, $M_4\ll m_{E4}$, under the assumption of $f_{Rk}\ll m_{E4}$ 
a block diagonalisation of  matrix (\ref{eq:EffNeutralFermionMassMatrix}) 
(i.e. decoupling the pseudo-Dirac pair $(\nu_4,\,N_4)$)  leads to
\begin{equation}
\label{eq:mNu}
m_\nu \simeq m^\mathrm{ss} + m^\mathrm{tree}
\end{equation}
with 
$m^\mathrm{ss} \equiv - m M^{-1} m^T$ 
and
\begin{equation}
m^\mathrm{tree}=\frac{1}{m_{E4}} \left[\left(m M^{-1}f_R
    f_L^T\right)+\left( ... \right)^T\right] . 
\end{equation}
The contribution $m^{tree}$ is linear in the light Dirac mass $m$ and therefore
can be considered as a new realisation of the linear see-saw~\cite{Barr:2003nn}.
Up to  high order corrections the total mass matrix in \Eqref{eq:mNu} can be rewritten as
\begin{equation}
 m_\nu =  - \left(m- \frac{1}{m_{E4}}f_L f_R^T\right)
 \left(M+\frac{M_{4}}{m_{E4}^2}f_R f_R^T\right)^{-1} 
 \left(m- \frac{1}{m_{E4}}f_L f_R^T\right)^T\;.
\end{equation}
Thus the total mass matrix  can be considered either as a combination of linear and ordinary 
type-I see-saw  (as in \Eqref{eq:mNu}) or as a type-I see-saw with a modified Dirac neutrino mass term. 

We can rewrite the contribution of the fourth generation to the
$\alpha\beta$ matrix element of $m_\nu$ as 
\begin{equation}
\label{eq:tree}
m^\mathrm{tree}_{\alpha\beta} \simeq 
m_{E4}\sum_k\frac{m_{\alpha k} }{M_{k}} (U_L)_{\beta 4} (U_R)_{k 4}+
(\alpha\leftrightarrow \beta)\;.
\end{equation}
Hence,  $m^\mathrm{tree}_{\alpha\beta}$ is  suppressed by the left- and right-handed
mixing in addition to the usual see-saw type factor. As the
leading contribution has rank one and the sub-leading ones are
suppressed, it can only generate one mass scale and the ordinary
see-saw contribution cannot be completely neglected.
We can compare the contributions of
the fourth generation via the $k^\mathrm{th}$ RH neutrino with the see-saw contribution of
the $k^\mathrm{th}$ RH neutrino to the light neutrino mass matrix as 
\begin{equation}
\frac{(m^\mathrm{tree}_k)_{\alpha\beta}}{(m^\mathrm{ss}_k)_{\alpha\beta}}\simeq
\frac{(U_R)_{k4} (U_{L})_{\alpha4}}{m_{k\beta }/m_{E4}}+
(\alpha\leftrightarrow \beta)\;.
\end{equation}
Therefore the tree level contributions via the $k^\mathrm{th}$ RH neutrino  dominates
over the seesaw if 
\begin{equation}
(U_R)_{k4} (U_{L})_{\alpha4}\gtrsim m_{k\beta }/m_{E4}. 
\end{equation}

\subsection{Two-Loop Mechanism \label{sec:Nu2loop}}

\begin{figure}[tb]
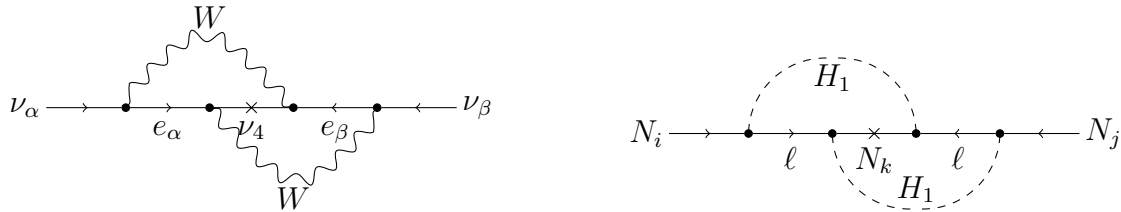

\subfigure[\,Two $W$ exchange contribution to light neutrino mass matrix.
\label{fig:TwoLoopNu}]{\FMDG{TwoLoopNuMass}}
\hspace{1cm}
\subfigure[\,Rank changing two loop diagram contributing to Majorana
mass term.\label{fig:2loopRHNuMass}]{\FMDG{RenormalizationOfRHMajoranaMass}}
\caption{The two loop diagrams which describe contributions to Majorana neutrino masses. 
Shown are the diagrams for the light neutrinos (left), 
and for the heavy RH neutrino (right).}
\end{figure}

If the components of the fourth generation neutrino are Majorana particles, they induce a
Majorana mass term for the light neutrinos at two
loop level, $m^\mathrm{loop}$, see \Figref{fig:TwoLoopNu}~\cite{Babu:1988ig}. 
We will consider the system of five neutrinos $(\nu_\alpha, \nu_E, N_4)$
after decoupling of the heavy RH neutrino components. 
The tree level mass matrix is then given by Eq. (\ref{eq:EffNeutralFermionMassMatrix}) 
and we neglect the see-saw contributions to the $\nu_\alpha -\nu_E$ as well as $\nu_E-\nu_E$ elements.  

The expression for the two-loop generated Majorana masses given in Eqs.~(\ref{eq:2loop},\ref{eq:2loopEV}) 
of \cite{Babu:1988ig} can be rewritten in the flavour basis as 
\begin{equation}
m^\mathrm{loop}_{A B} \simeq - \frac{g^4}{m_W^{4}} m_{E 4}^2 M_{4}  m_A^2 m_B^2
(U_L)_{A 4}(U_L)_{B 4} I_{A B}\;. 
\label{2loop}
\end{equation}
Here the indices $A, B = e, \mu, \tau, E$ run over four generations,  $m_{A}$ and $m_{B}$ are  
the charged lepton masses,  
$g$ is  the SU(2) gauge coupling and $m_W$ is the mass of the
$W$-boson. The integral $I_{A B}$ equals 
\begin{multline}
I_{A B} = \int \frac{\dd^4 p}{(2\pi)^4}\int\frac{\dd^4 q}{(2\pi)^4} \frac{p\cdot 
q}{(p^2-m_A^2)(q^2-m_B^2)}\frac{1}{(p+q)^2-m_{N1}^2}\frac{1}{(p+q)^2-m_{N2}^2}\\
\left[\frac{1}{p^2q^2}-\frac34\frac{1}{p^2-m_W^2}\frac{1}{q^2-m_W^2}\right]\;.
\end{multline}
Here  $m_{N1,2}$ are the eigenvalues of the mass matrix of the fourth generation neutrino states.
According to \Eqref{2loop},  the two loop generated masses  depend on the mixing of the 
fourth generation neutrino with the light neutrinos, $U_{A 4}$, 
the Dirac mass of the fourth generation 
neutrino, $m_{E4}$,  and the Majorana mass of the RH neutrino, $M_4$.  

There are two different two-loop contributions to the mass matrix of light neutrinos:  
(i) the direct one which follows from \Eqref{2loop} for $A, B = e, \mu, \tau$, and (ii) 
the contribution via the ${EE}$-element, the Majorana mass of $\nu_E$, $m_{EE}$ generated 
in 2 loops. As we will see, the latter dominates due to hierarchy of the charged 
lepton masses: $m_E \gg m_{e,\mu,\tau}$. In fact, this contribution can be computed  
in the approximation of  vanishing charged lepton masses $m_{e,\mu,\tau} = 0$.  

Let us compute the second contribution in the pseudo-Dirac case    
when the masses of neutrinos of the fourth generation equal $m_{N1}\simeq m_{N2}\simeq m_{4}$  
and the splitting between them given by $M_4$ is small: $M_4 \ll m_4$. 
According to \Eqref{2loop} the  $m_{EE}$ element is given by 
\begin{equation}
m_{EE}^\mathrm{loop} \simeq -g^4 m_W^{-4}m_{4}^2 M_{4} m_E^4
(U_L)_{E 4}^2 I_{EE}\;.
\end{equation}
In the limit $m_W\ll m_4\ll m_E$ the integral $I_{EE}$  
equals approximately~\footnote{See the appendix 
of \cite{Aparici:2011nu} for the evaluation of this integral.}
\begin{equation}
\label{eq:IEE}
I_{EE}\simeq\frac{1}{(4\pi)^4 m_E^2} \left(\frac{\pi^2}{3}-2+\ln \frac{m_4^2}{m_E^2}\right)\;.
\end{equation}
Now we have the mass matrix (\ref{eq:EffNeutralFermionMassMatrix}) with non-zero elements in 
the fifth row and column and  non-zero  $EE$-element. 
Decoupling of the fourth (pseudo-Dirac) neutrino ({\it i.e.}~the see-saw diagonalisation with  
$\nu_E, N_4$  heavy block) contributes to the  masses of the light neutrinos:  
\begin{equation}
m^\mathrm{loop}  \simeq  m_{EE}^\mathrm{loop} \frac{f_L^T f_L}{m_{E4}^2- m_{EE}^\mathrm{loop} M_4}\; ,
\end{equation}
which in the  case $m_{E4}^2 \gg m_{EE}^\mathrm{loop} M_4$ leads  to 
\begin{equation}
\label{eq:2loopScale}
m^\mathrm{loop}_{\alpha \beta}  \simeq  m_{EE}^\mathrm{loop} 
\frac{f_{L \alpha}}{m_{E4}} \frac{f_{L \beta}}{m_{E4}} = 
m_{EE}^\mathrm{loop}  (U_{L})_{\alpha 4} (U_{L})_{\beta 4}
\;.
\end{equation}
The resulting contribution to the light neutrino mass
matrix is
\begin{eqnarray}
\label{eq:2loop}
m_{\alpha\beta}^\mathrm{loop}&\simeq&-\frac{g^4m_E^4}{m_W^4}M_4 m_4^2 I_{EE} \left(U_L\right)_{E4}^2\left(U_L\right)_{\alpha4}\left(U_L\right)_{\beta4}\nonumber\\
&=&-\frac{g^4\left(U_L\right)_{E4}^2}{(4\pi)^4}\left(\frac{\pi^2}{3}-2+\ln \frac{m_4^2}{m_E^2}\right)\frac{m_4^2 m_E^2}{m_W^4} M_4 \left(U_L\right)_{\alpha4}\left(U_L\right)_{\beta4}\;.
\end{eqnarray}
The mass matrix formed by the loop contribution via $m_{EE}^\mathrm{loop}$ (\ref{eq:2loop})
is singular (rank 1) with the unique non-zero eigenvalue 
\begin{equation}
\label{eq:2loopEV}
m^\mathrm{loop}_4 \approx \frac{g^4 m_E^4}{ m_W^{4}} M_{4} m_4^2 \left| I_{EE} \sum_\alpha (U_{L})_{\alpha 4}^2\right|\;.  
\end{equation}
When the masses of three SM charged lepton  are taken into account, the two
loop contribution obtains full rank. 
However  due to  a strong  hierarchy
of these masses the two loop contribution cannot explain neutrino masses by themselves~\cite{Aparici:2011nu}. 
The contribution (\ref{eq:2loop}) dominates over the direct two loop contribution 
(\ref{2loop}) due to three known SM leptons. Indeed, the ratio of the two  equals: 
\be
\frac{m_\alpha^2 m_\beta^2}{m_E^4} \frac{I_{\alpha \beta}}{I_{EE}}. 
\ee
The flavour structure of the loop contribution is determined by 
the mixing matrix elements $(U_{L})_{\alpha 4}$. 
Note that the results  considered here can be immediately obtained 
by computing diagrams with 
the would-be Goldstone bosons, then all the substantial quantities arise from vertices.

In the see-saw case for the fourth generation masses, $M_4 \gg m_4$ the two-loop contribution is 
suppressed by $f_{L\alpha}/m_{E4}$ only,  because the direct Majorana mass term $M_4$ is
large. Consequently, the mixing between the fourth generation and the three
light SM generation has to be small.

\subsection{Radiative Generation of Fourth Generation Singlet Majorana
  Mass\label{sec:RadGenMajorana}}

Similarly to $m_{EE}$ considered in the previous subsection, the  RH Majorana neutrino mass 
$M_4$ can be generated 
at the two loop level. The relevant (rank changing) two loop diagram is shown in
\Figref{fig:2loopRHNuMass},  which results
in the following expression for the mass in the \MSbar\ renormalisation scheme
\begin{equation}
M_{ij}^\mathrm{loop} = \frac{2}{(16\pi^2)^2} \sum_{k=1}^4(Y^\dagger Y)_{ik}
\left(Y^\dagger Y\right)_{jk} M_{k}
\left(\frac1\epsilon+\frac12+\ln\frac{\mu^2}{M_{k}^2}\right)\; . 
\end{equation}
Here $i,j=1,\dots,4$ and the Higgs mass has been neglected.
The RG produced fourth generation Majorana mass can be estimated (neglecting diagrams 
with light charged fermions) as 
\begin{equation}
\label{eq:M4TwoLoop}
M_{4}^\mathrm{loop} \simeq \frac{y_{4}^4}{(8\pi^2)^2} 
\sum_{i=1}^3\left[(U_{R})_{i4}^* 
(U_{R})_{E4}\right]^{2} M_{i} \ln \frac{M_{i}}{\Lambda}
\end{equation}
with $\Lambda$ being the high scale, at which the theory is defined,
and $y_{4} = m_{E4}/\vEWnu$ being the neutrino 
Yukawa coupling of the fourth  generation, 
which dominates over the other Yukawa couplings. An estimate can also been be found in \cite{Aparici:2011eb}.
In addition to the RG running, the finite part of the counter term
leads to a scheme dependent
threshold correction~\footnote{Notice that this result 
for the radiatively induced RH neutrino mass is 
rather general and valid beyond the four generation context. It is particularly interesting
in case of the nearly singular see-saw when one of the mass eigenvalues is
considerably smaller than the remaining ones.}. 
The main contribution comes from the diagrams with the charged lepton $E$. 
According to \Eqref{eq:M4TwoLoop}  $M_{4}^\mathrm{loop} \propto m_4^4 U_R^2$ and 
therefore it quickly decreases with $m_4$. For $U_R \sim U_L$ inspired by the L-R symmetry and a single Higgs, {\it i.e.}~$H_1=H$ and $H_2=H^C$,
the mass can be estimated as $M_4=1.0\,\GeV$ for $U_R=0.001$, $m_4=400\,\GeV$, $M_i=10^8\,\GeV$ and $\Lambda=10\,M_i$.

Neglecting an accidental cancellation, we expect that the
fourth generation Majorana mass, $M_{4}$,  has at least the size
of the radiatively generated contribution given in
\Eqref{eq:M4TwoLoop}:  $M_{4} \geq M_4^\mathrm{loop}$ . Using $M_4^\mathrm{loop}$ only, 
we can estimate the size of the
contributions to the neutrino masses using \Eqref{eq:2loop}: 
\begin{align}
m_{\alpha\beta}^\mathrm{loop}&\simeq
-\frac{y_4^4 g^4 \left(U_L\right)_{E4}^2}{4 (8\pi^2)^4} \left(\frac{\pi^2}{3}-2+\ln \frac{m_4^2}{m_E^2}\right)\frac{m_4^2 m_E^2}{m_W^4}  \left(U_L\right)_{\alpha 4} \left(U_L\right)_{\beta4}
\sum_{i=1}^3\left[\left(U_R\right)_{i4}^*\left(U_R\right)_{E4}\right]^2 M_i \ln \frac{M_i}{\Lambda}\;.
\label{eq:boundTauTau}
\end{align}
Effectively it is generated at four loops level. 
According to \Eqref{eq:boundTauTau}, and since $(U_L)_{E4} \approx (U_R)_{E4} \approx 1$   
the  bound on the neutrino mass scale 
leads to  an upper bound on the combination 
$|(U_{R})_{i4} (U_{L})_{\alpha 4}|^2  M_{i}$.  
The bound strongly depends on $m_4 = y_4 \vEWnu$ and $m_E$.

\begin{figure}[tb]
\begin{center}
\includegraphics[width=8cm]{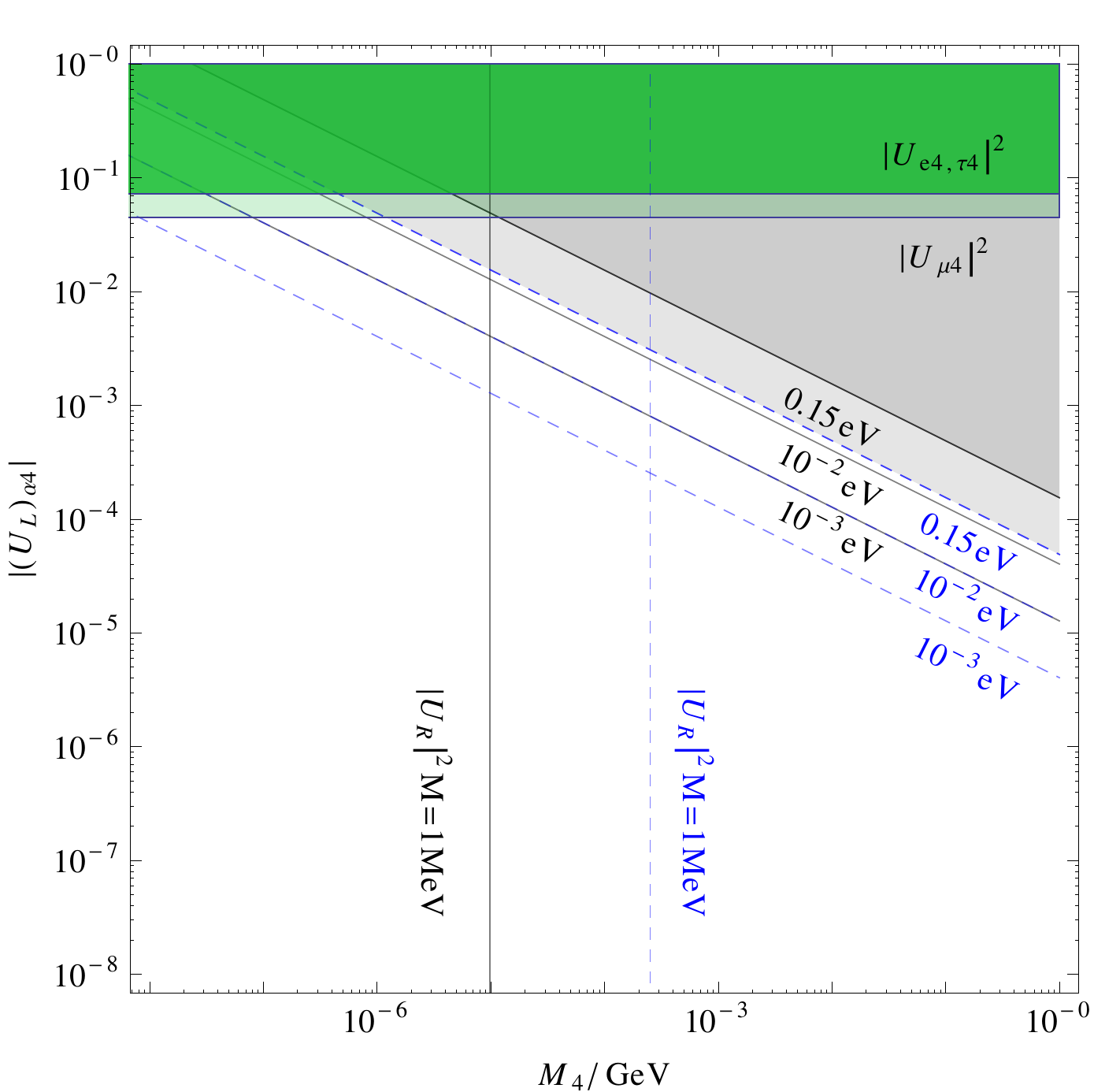}
\caption{Iso-contours of the two loop contribution 
to light neutrino masses in the $(U_L)_{\alpha 4} - M_4$ plane 
for different values of the Dirac neutrino
  mass, $m_4$, and charged lepton mass, $m_E$. The black solid lines correspond to 
  $m_4=400\,\GeV$ and $m_E=600\,\GeV$, while the blue dashed lines
  correspond to $m_4=900\,\GeV$ and $m_E=1\,\TeV$. The shaded region is excluded by the cosmological bounds on the
  neutrino mass. The green shaded areas are
  excluded by the bounds on the mixing angles $U_{\alpha4}$ (see 
  \Secref{sec:bounds}). The vertical lines correspond to
  the Majorana mass $M_4$  induced at the two loops level 
  for a RH neutrino mixing $|U_R|=10^{-4}$ and
  RH neutrino mass $M=100\,\TeV$ with  the cutoff scale~$\Lambda=1000\,\TeV$.
\label{fig:Exclusion}}
\end{center}
\end{figure}

\Figref{fig:Exclusion} shows the iso-contours  of the two loop
contribution to the light neutrino masses in the plane of mixing angle $|U_{\alpha4}|$ 
and the fourth generation Majorana mass $M_4$. 
The equation for these contours is given by \Eqref{eq:2loop} 
which can be rewritten as 
\be
\left|\left(U_L\right)_{\alpha4}\right|=\sqrt{\left|\frac{m^\mathrm{loop}}{C_2M_4}\right|}\frac{m_W^2}{m_E m_4}\;,
\ee 
where 
\be
C_2\equiv -\frac{g^4\left(U_L\right)_{E4}^2}{(4\pi)^4}\left(\frac{\pi^2}{3}-2+\ln\frac{m_4^2}{m_E^2}\right)\;.
\ee
The vertical lines in \Figref{fig:Exclusion} indicate the
Majorana mass $M_4$ generated by the two loop correction (\ref{eq:M4TwoLoop}) 
for $y_4 = m_4/\vEWnu$. 
So, $|(U_L)_{\alpha 4}| \propto 1/ \sqrt{M_4}$, and furthermore the mixing becomes small with 
increase of $m_E$. Values of mixing parameters $(U_L)_{\alpha 4}$ at the level achievable by 
the direct searches can be obtained only for very small 
RH neutrino masses: $M_4 < 1$ keV. On the other hand  $M_4$  
of the size $\sim 1$ GeV  requires $(U_L)_{\alpha 4} \sim 10^{-5}$.

The bound on the light neutrino mass scale strongly constrains the combination
$|(U_{L})_{\alpha4}^2\,M_4|$. Barring accidental cancellations, we
expect the fourth generation RH neutrino mass to be at least of the
scale generated by the two loop diagrams.  Under this
assumption, the mixing parameter  $|U_{\alpha4}|$ should  be
smaller than the value at the intersection of the vertical line 
with the lower border of the excluded
shaded area.

These two loop corrections do not exist in SUSY due to the non-renormalisation theorem for the superpotential~\cite{Grisaru:1979wc}. However, if there are RH neutrinos with a mass below the SUSY breaking scale, there are quantum corrections. The larger particle content in the SUSY version compensates for the smaller logarithms coming from the RG corrections (see {\it e.g.} \cite{Davidson:2006tg,*Ray:2010fa}~for two loop corrections to the light neutrino mass matrix).

\section{Phenomenology of Fourth Generation Neutrinos}
\label{sec:discussion}

\subsection{Existing Bounds on Fourth Generation Leptons}
\label{sec:bounds}

Let us summarise the bounds on masses and mixing of the fourth generation particles which we will use 
in our analysis. The collider searches give~\cite{Nakamura:2010zzi}
\begin{equation}
\label{eq:limitsEN}
m_E >100.8 \,\GeV\, , ~~~ m_N > (80.5 - 101.5)\,\GeV\,.
\end{equation}
The range of values for the lower bounds on $m_N$ in \Eqref{eq:limitsEN} 
originates from different 
search channels, $N\rightarrow W^* + (e,\,\mu,\,\tau)$ under the assumption of a  100\% branching ratio  
in a given channel. The bounds depend also on the nature of the neutrino: 
for Majorana neutrino they are  about  $10\,\GeV$
weaker than for Dirac neutrinos. 
The bounds rely on the assumption that only one heavy neutral lepton can be produced.
A recent reanalysis~\cite{Carpenter:2010dt} shows that the bounds can be relaxed
when two heavy neutral leptons are accessible, like in the framework of
pseudo-Dirac neutrinos. Under the assumption of a mass splitting
$M_{N_2}-M_{N_1} > 10\,\GeV$
between the two heavy neutral states $N_{1,2}$, the study of $e^+e^-\rightarrow Z^*\rightarrow N_i N_j \rightarrow l
W^* l W^* Z^{*0,1,2}$ leads to the
bounds $62.1\,\GeV (W^*\tau)$, $79.9\,\GeV (W^*\mu)$ and $81.8\, \GeV
(W^* e)$. The number of $Z^*$-bosons depends on the number of
produced $N_2$ via $N_2\rightarrow N_1 Z^*$. In our study, we mainly consider smaller mass
splittings $M_{N_2}-M_{N_1}$, where the branching ratio of
$N_2\rightarrow l W^*$ dominates over the one of $N_2\rightarrow N_1
Z^*$. This leads to an interference between $N_{1,2}\rightarrow lW^*$ and we expect the bounds to become weaker.

A study of the sensitivity of the Tevatron to a fourth generation
neutrino~\cite{Rajaraman:2010wk} shows that it can put a
lower bound $m_N> 175\,\GeV$ and has a $3\sigma$ discovery
potential for $m_N <150\,\GeV$ with 5 fb$^{-1}$.
The LHC can exclude fourth generation charged leptons up
to $250\,\GeV$~\cite{Carpenter:2010bs,*Carpenter:2010sm}.

The electroweak precision tests (which include quark mixing but neglect leptonic mixing)  
constrain the mass splitting between the fourth generation
leptons~\cite{Eberhardt:2010bm} 
\begin{equation}
|m_E - m_{N} | < 140\,\GeV  , 
\end{equation}
indicating that  the masses of the fourth generation leptons should be of the same order of 
magnitude.

The leptonic mixing angles are constrained by searches for the
radiative $\mu^-$ and $\tau^-$ decays, $\ell_i\rightarrow\ell_j\gamma$, as
well as by kaon and pion decays. The
limits given in~\cite{Lacker:2010zz,*Menzel:2011ff} read
\begin{equation}
\label{eq:PMNS}
 U_\mathrm{PMNS}=\left(\begin{array}{cccc}
                  . & . & . &<0.073 \\
 . & . & . &<0.045 \\
 . & . & . &<0.072 \\
<0.092&<0.092&<0.092&>0.9958
                \end{array}
\right)\;.
\end{equation}
There is an even stronger bound on $U_{\mu4}^* U_{e4}$ from the $\mu - e$ conversion: $|U_{\mu4}^* U_{e4}|<0.4\cdot10^{-4}$ for  $m_N > 100$ GeV~\cite{Deshpande:2011uv}.

The influence of mixing of light generations on the masses of the fourth neutrino
can be neglected.  We can  estimate the maximal allowed 
value of $M_{4}$ which is realized in the  see-saw limit, $M_{4}\gg m_4$, as  
$M_{4} =  m_4^2/ m_N$. Using 
the unitarity upper limit on 
$m_{4}\lesssim 1.2\, \TeV$~\cite{Chanowitz:1978uj,*Chanowitz:1978mv} 
and the LEP exclusion limit for an additional neutral lepton 
$m_N\sim\,100\,\GeV$ we find  
\be 
M_{4} 
\lesssim 14\,\TeV .
\ee

\subsection{Neutrinoless double beta decay and Cosmological Bounds}
\label{sec:0Nu2Beta}

All three main mechanisms of light neutrino mass generation are essentially 
of the see-saw type and 
the $\beta\beta_{0\nu}$-decay
proceeds via  the neutrino exchange  only.
Therefore, we can apply here the results of~\cite{Blennow:2010th}. 
Following~\cite{Blennow:2010th}, we separate  the contributions 
to the amplitude of the decay from  a
heavy mass eigenstates with a mass $m_I\gg m_\pi\sim100\,\MeV$, and from the
light neutrino mass eigenstates with mass $m_i\ll m_\pi$: 
\begin{equation}
\label{eq:nu2BetaAmp}
A \propto \sum_i^\mathrm{light} m_i U_{ei}^2 M^{0\nu\beta\beta}(m_i)
+\sum_I^\mathrm{heavy} m_I U_{eI}^2 M^{0\nu\beta\beta}(m_I)\;,
\end{equation}
where the masses $m_{i,I}$ and mixing angles $U_{ei}$, $U_{eI}$ are
defined by
\begin{equation}
\label{eq:mU}
U^* \diag(m_1,\dots,m_8)U^\dagger=\mathcal{M}
\end{equation}
with $\mathcal{M}$ being the $8\times8$ neutral fermion mass matrix.

The two loop direct contribution to the $m_{ee}$ element in the flavour
basis is negligible being proportional to $m_e^4$. 
Therefore, according to  \Eqref{eq:NeutralFermion} $m_{ee} \approx 0$. 
In terms mixing angles and mass 
eigenstates defined in \Eqref{eq:mU} this condition can be expressed as    
\begin{equation}
\label{eq:mUeRel}
\sum_i^\mathrm{light}m_i U_{ei}^2 + \sum_I^\mathrm{heavy} m_I U_{eI}^2 \approx0\;.
\end{equation}

The nuclear matrix elements $M^{0\nu\beta\beta}$ in \Eqref{eq:nu2BetaAmp} 
include neutrino propagators: 
\begin{equation}
D_\nu \propto \left\{ 
\begin{array}{lll}
\frac{1}{p^2 - m_i^2} \approx \frac{1}{p^2}, &  {\rm for} &  m_i \ll m_\pi\\ 
\frac{1}{p^2 - m_I^2} \approx - \frac{1}{m_I^2}, &  {\rm for} &  m_I \gg m_\pi 
\end{array}
\right. , 
\end{equation}
where $m_\pi \sim 1/r_N$ is the pion mass, which gives the inverse size of the nucleus radius,  
$r_N$. Therefore  $M^{0\nu\beta\beta}$ practically does not depend 
on the mass of the exchanged light neutrinos:  
$M^{0\nu\beta\beta}(m_i)\approx M^{0\nu\beta\beta}(0)$.  
For heavy neutrinos the matrix element  decreases 
as  $M^{0\nu\beta\beta}(m_I)\propto m_I^{-2}$.  Consequently, 
the ratio of the matrix elements
\be
\frac{M^{0\nu\beta\beta}(m_I)}{M^{0\nu\beta\beta}(m_i)} 
\sim \frac{m_\pi^2}{m_I^{2}} \ll 1 . 
\ee
Using relation (\ref{eq:mUeRel}) 
we can rewrite the amplitude   \Eqref{eq:nu2BetaAmp} in
the following way 
\begin{align}
A &\propto \sum_i^\mathrm{light} m_i U_{ei}^2 M^{0\nu\beta\beta}(0)
+\sum_I^\mathrm{heavy} m_I U_{eI}^2 M^{0\nu\beta\beta}(m_I)  \nonumber\\
&= \sum_I^\mathrm{heavy} m_I U_{eI}^2 (M^{0\nu\beta\beta}(m_I)-M^{0\nu\beta\beta}(0)) 
\nonumber\\
&\approx  -\sum_I^\mathrm{heavy} m_I U_{eI}^2 M^{0\nu\beta\beta}(0) 
= \sum_i^\mathrm{light} m_i U_{ei}^2 M^{0\nu\beta\beta}(0) 
\; , 
\end{align}
where we used $ M^{0\nu\beta\beta}(m_I)\ll M^{0\nu\beta\beta}(0) \approx
M^{0\nu\beta\beta}(m_i)$ in the first and third line and
\Eqref{eq:mUeRel} in lines two and three. Hence, the dominant contribution to
$\beta\beta_{0\nu}$-decay is from light neutrinos.

Thus,  the bound from $\beta\beta_{0\nu}$-decay is reduced to 
the bound on the effective Majorana mass of the electron neutrino due to light neutrinos only. 
That is,  the $\beta\beta_{0\nu}$-decay gives a bound on the light neutrino masses, 
which has also been pointed out in \cite{Aparici:2011nu}, and as far as this bound is satisfied, no other bounds on the model appear.  
Hence, $\beta\beta_{0\nu}$-decay restricts the model  via the light masses only. 

Notice that in the discussion of $\beta\beta_{0\nu}$-decay 
in~\cite{Lenz:2011gd,*Lenz:2010ha}, the light contribution has been
neglected. 

At the moment, cosmology gives even a stronger bound on light neutrino masses than the 
$\beta\beta_{0\nu}$-decay. 
We  took $m_0\lesssim0.15\,\eV$ for an individual neutrino as reference
value which originates from the bound  
$\sum m_i\lesssim 0.44~\eV$~\cite{Hannestad:2010yi}.

\subsection{Comparison of Different Contributions to Neutrino Mass}

As we have found in the previous section, in models with four families of  fermions, 
generically there are three contributions to the light neutrino masses 
from three different mechanisms: (i) the usual see-saw type-I,  $m^\mathrm{ss}$;  
(ii) the tree-level contribution  $m^\mathrm{tree}$ due to mixing of the light neutrinos with $\nu_4$ is  essentially another see-saw,  it is linear in the usual Dirac mass
matrix; (iii) the 2-loop contribution induced by
the Majorana mass term of the neutrino of the fourth family,
$m^\mathrm{loop}$. These three contributions have different flavour structures but partially 
correlate. For a given Dirac mass matrix of light neutrinos:  
 $m^\mathrm{ss} = m^\mathrm{ss} (M_k)$,
$m^\mathrm{tree} =  m^\mathrm{tree} (M_k, U_{R}, U_L)
$ (\Eqref{eq:tree}), and  $m^\mathrm{loop} = m^\mathrm{loop} (M_k, U_{R}, U_L)$ 
(\Eqref{eq:2loop}). 

In what follows, we will  consider these contributions in the 
case of  a vanishing fourth generation
Majorana mass term $M_{4}$ at tree-level. The mass $M_4$ is constrained 
on the one hand by the invisible $Z$-decay 
width and on the other hand by the bound on the neutrino mass which is
induced at two loop (see \Figref{fig:Exclusion}). Furthermore, 
we restrict ourselves to a single Higgs doublet, {\it i.e.}~$H_1=H$  and $H_2=H^C$. Similar conclusions apply in a two Higgs doublet model. The main difference is an increased  neutrino Yukawa coupling, for fixed 
values of masses. This leads to a larger loop contribution to neutrino masses.
The following discussion does not depend on the Higgs mass $m_H$, as long as it is negligible compared to the heavy RH neutrino masses $M_i$, $i=1,2,3$. If they are of a similar magnitude, the expression (\ref{eq:M4TwoLoop}) for $M_4$ will change, but it remains valid 
as an order of magnitude estimate. Hence, the following conclusions are also valid for a heavy SM Higgs $m_H\gtrsim 600\,\GeV$.
 
We will first discuss the ``1+1''  generation case: one light generation and the fourth generation~\footnote{The results can be directly applied to one
specific matrix element in case of 3+1 generations.}.
In this case we have one light neutrino with Dirac mass 
$m \equiv m_{\alpha i}$  and one very heavy RH neutrino with mass 
$M \equiv M_i$. We can introduce a single parameter which characterises mixing of the light neutrino 
with the neutrino of the fourth generation:   
\begin{equation}
\xi \equiv  (U_{L})_{\alpha 4} (U_R)_{i4} . 
\end{equation}
In terms of this parameter the tree level contribution  (\ref{eq:tree}) can be written as 
\begin{equation}
\label{eq:treem}
m^\mathrm{tree} = m_{E4} \frac{2m}{M} \xi .  
\end{equation}
The loop contribution (\ref{eq:boundTauTau}) is then 
\begin{equation}
\label{eq:loopm}
m^\mathrm{loop}=C^\mathrm{loop} m_E^2 m_4^6 \xi^2 M \ln \frac{M}{\Lambda}\;,
\end{equation}
where 
\be 
C^\mathrm{loop}\equiv-\frac{g^4}{4 (8 \pi^2)^4 m_W^4} \frac{1}{{v_\mathrm{EW}^\nu}^4}\left(\frac{\pi^2}{3}-2+\ln\frac{m_4^2}{m_E^2}\right)
\ee
and we have taken into account that $(U_L)_{E4} \approx (U_R)_{E4} \approx 1$. 
Let us underline that mixing parameters of the fourth neutrino enter the contributions 
only in the combination $\xi$. The other relevant parameters are $m_E$,  $m_4$, $M$ and $m$. 
Relative contributions of different mechanisms depend on values of these parameters. 

\begin{figure}[tb]
\begin{center}
\subfigure[$m=31\,\MeV$\label{fig:con2mMeV}]{\includegraphics[width=8cm]{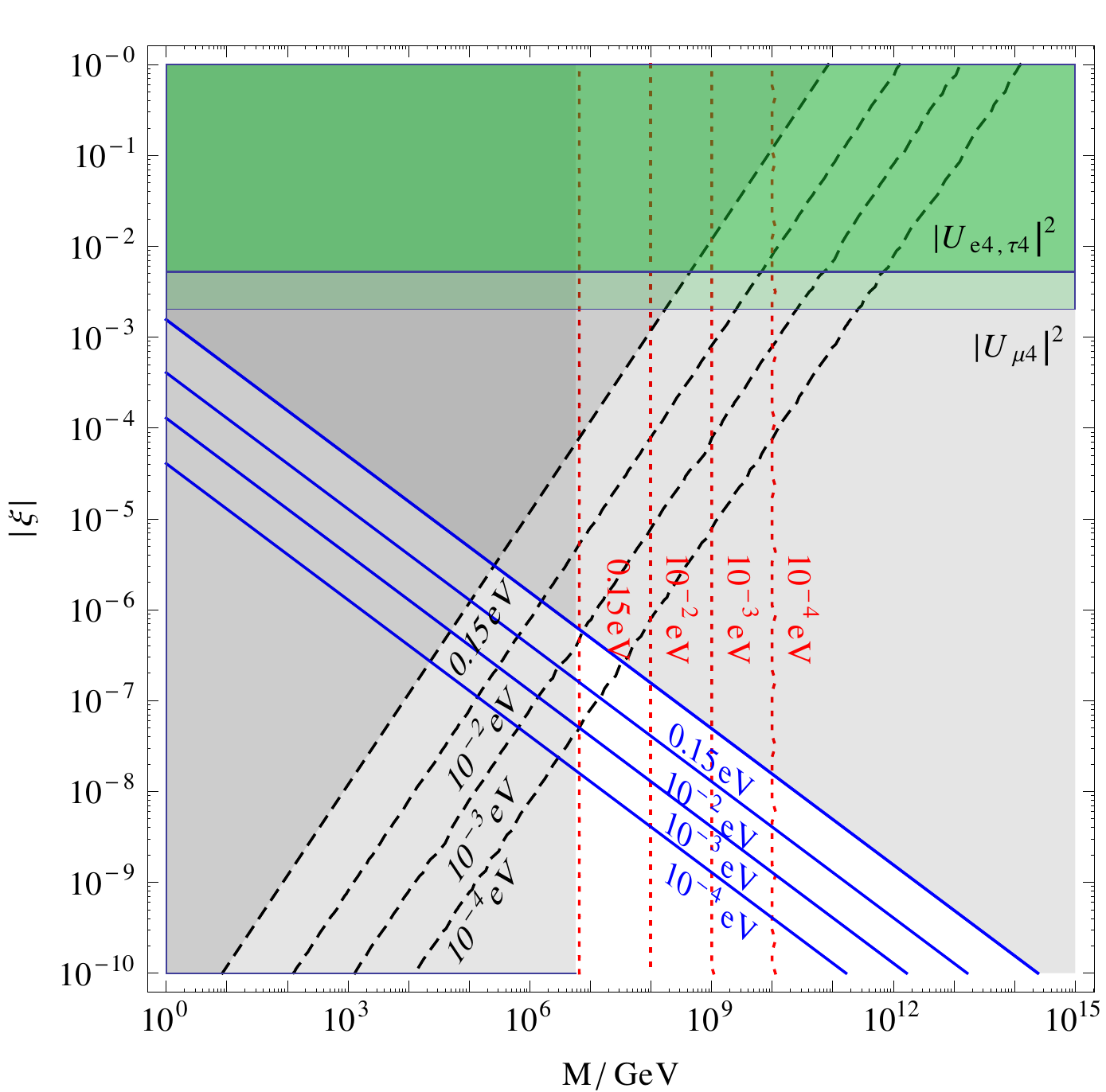}}
\hspace{0.1cm}
\subfigure[$m=m_e$\label{fig:con2mGeV}]{\includegraphics[width=8cm]{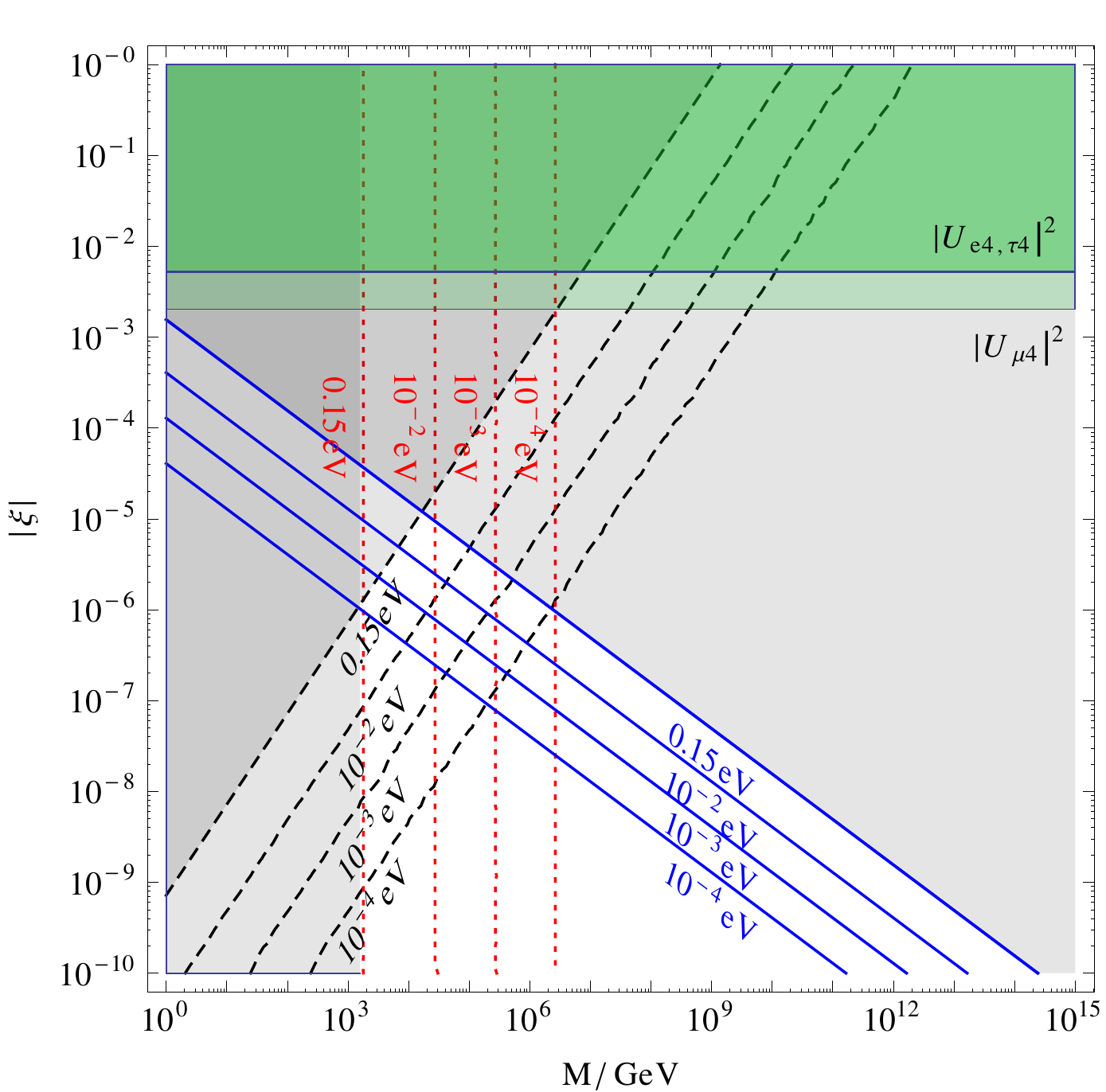}}
\caption{Iso-contours of different contributions to light neutrino masses (numbers at the curves) from different mechanisms in the $\xi- M$ plane for 
$m_4=400\,\GeV$ and $m_E=600\,\GeV$ and two different values of the Dirac mass $m$. Black dashed lines correspond to the tree-level contribution, blue solid lines to the loop contribution and red dotted ones to the usual see-saw contribution. Note that the Higgs mass $m_H$ has been neglected in the calculation of the radiatively induced $M_4$. Hence for $M\sim m_H$, the loop contribution can only be considered as an order of magnitude estimate.
\label{fig:con2}}
\end{center}
\end{figure}
In \Figref{fig:con2}  we show the iso-contours of different contributions to neutrino mass 
in the $\xi - M$ plane for fixed values of 
$m_E$,  $m_4$ and $m$.  The equations for these contours can be readily obtained 
from \Eqref{eq:treem}  and \Eqref{eq:loopm}. 
For a given value of  
$m^\mathrm{tree}$ we find from \Eqref{eq:treem} the following dependence of 
$\xi$ on $M$:  
\be 
\label{eq:xitree}
\xi^\mathrm{tree} = M \frac{m^\mathrm{tree}}{2m m_{E4}}. 
\ee
That is,  $\xi^\mathrm{tree}$  linearly increases with $M$; it is proportional to  
$m^\mathrm{tree}$ and inversely proportional to  $m$. The iso-contours of the tree-level contribution correspond to the black dashed lines.

From \Eqref{eq:loopm} we obtain the analytic expression for iso-contours 
of the loop  contribution (blue solid lines in Fig. \ref{fig:con2}):  
\begin{equation} 
\label{eq:xiloop}
\xi^\mathrm{loop}=\left[\frac{m^\mathrm{loop}}{C^\mathrm{loop}m_E^2m_4^6}\right]^{1/2}\cdot \frac{1}{\sqrt{M\ln\frac{M}{\Lambda}}}\;.
\end{equation}
For $M \propto \Lambda$ this equation gives $\xi^\mathrm{loop} \propto 1/\sqrt{M}$. 

The usual see-saw contribution,  $m^\mathrm{ss} = - m^2/M$,  
does not depend on $\xi$ and the corresponding  iso-contours are just vertical 
red dotted lines in the  plot of \Figref{fig:con2}. 
According to \Figref{fig:con2},  the loop contribution dominated for large values of $M$ and 
small values of  $\xi$. The tree level contribution is larger for small $M$  and 
large $\xi$, whereas the usual see-saw dominates in the range of small $M$. 
For $m = 31\,\MeV$ (see \Figref{fig:con2mMeV}) the allowed region is  $M \gtrsim 10^{7}$ GeV and 
$\xi \lesssim 10^{-6}$. The tree-level contribution is negligible in this region and 
the total neutrino mass is determined by an interplay of the usual see-saw and 
the loop contributions. Furthermore, the see-saw dominates at smaller $M$ and $\xi$. 

With decrease of $m$, the relative contributions of  different mechanisms change: 
the  iso-contours of $m^\mathrm{loop}$ do not move,  the see-saw lines shift to smaller $M$ 
as $M \propto m^2$, whereas the iso-contours of  tree level contribution shift  as $M \propto m$, {\it i.e.}~weaker.  
Therefore, the tree-level contribution becomes important and can dominate for 
a small Dirac mass $m$ in the range of a small RH Majorana mass $M$ and relatively large $\xi$. 
  With increase of $m_4$,  $\xi^\mathrm{loop} \propto 1/m_4^2$ decreases faster than 
$\xi^\mathrm{tree} \propto 1/m_4$. Therefore the tree level contribution 
becomes substantial and the allowed region shifts to smaller $\xi$.  
Also with increase of the charged lepton mass $m_E$ the loop contribution 
increases. 

\begin{figure}[tb]
\begin{center}
\subfigure[\,$\xi=10^{-7}$\label{fig:con3}]{\includegraphics[width=8cm]{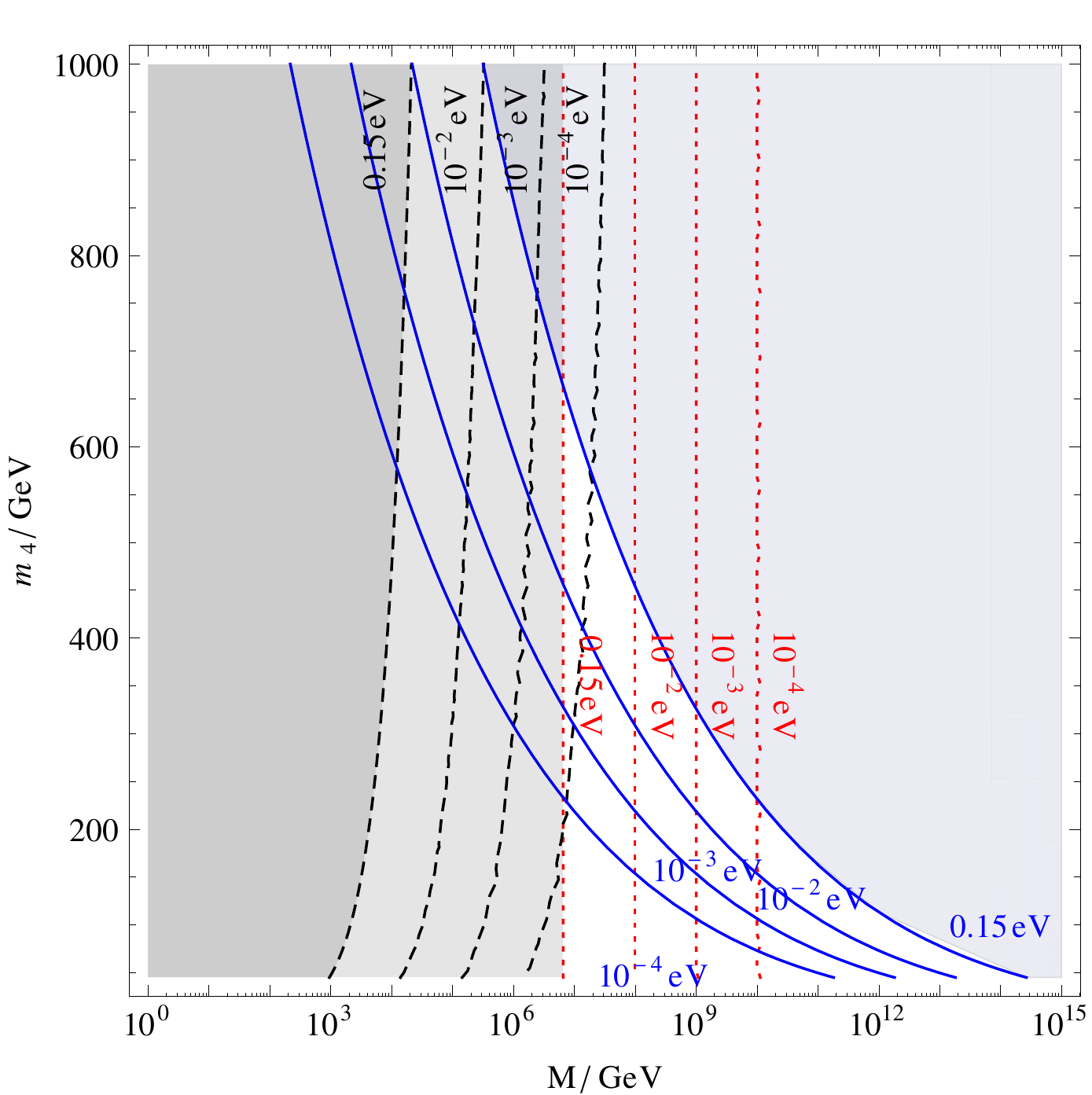}}
\hspace{0.1cm}
\subfigure[\,$\xi=10^{-9}$\label{fig:con4}]{\includegraphics[width=8cm]{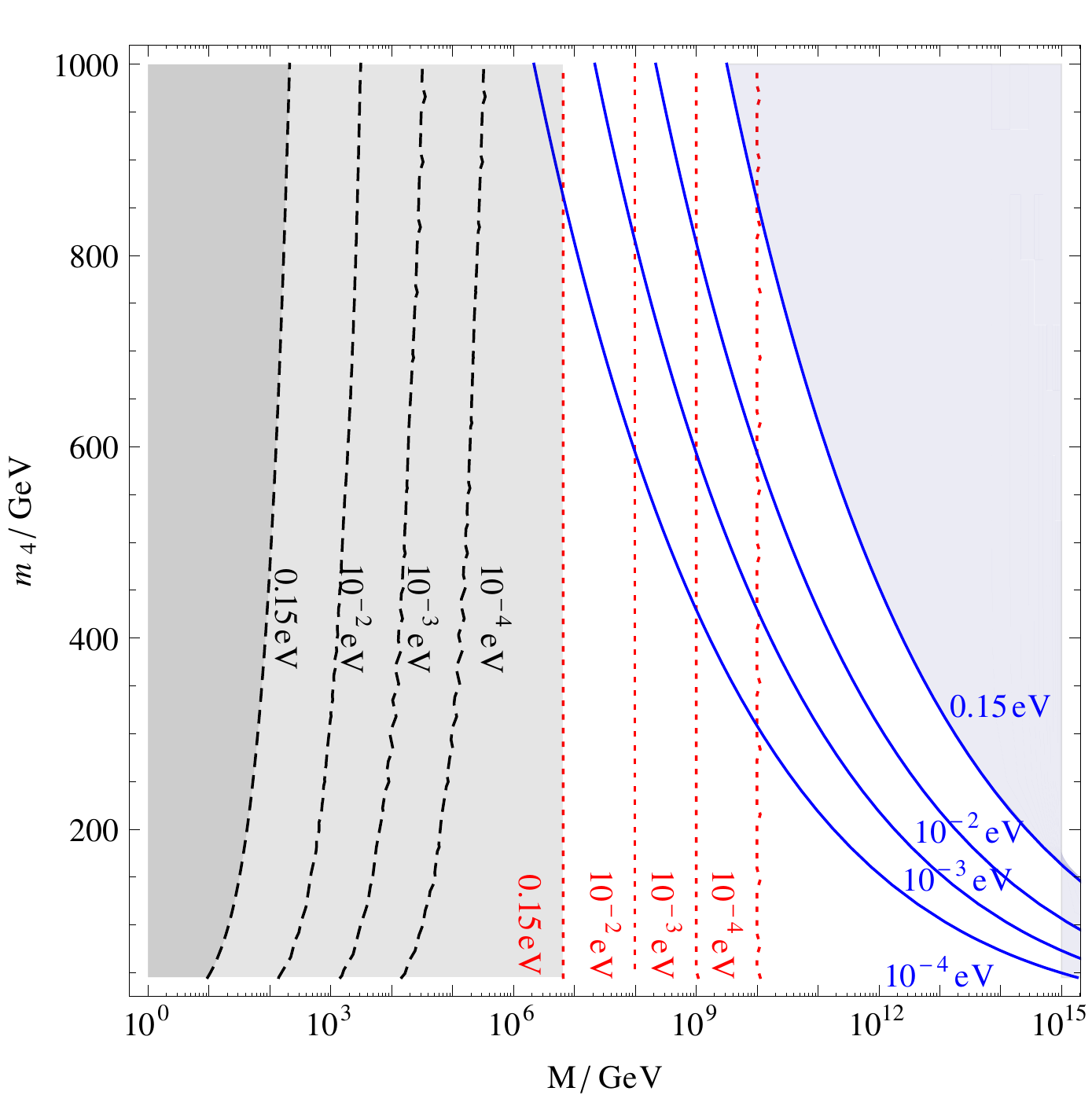}}
\caption{Iso-contours of the contributions to light neutrino masses (numbers at the curves) 
from different mechanisms in the $m_4 - M$ plane for
two different values of 
the mixing parameter $\xi$. The Dirac mass $m$ is fixed to $m=31\,\MeV$ and the fourth generation charged
lepton mass $m_E$ is fixed to be $200\,\GeV$ larger than $m_4$: $m_E  = m_4 + 200$ GeV. Black dashed lines correspond to the tree-level contribution, blue solid lines to the loop contribution and red dotted ones to the usual see-saw contribution.  Note that the Higgs mass $m_H$ has been neglected in the calculation of the radiatively induced $M_4$. Hence for $M\sim m_H$, the loop contribution can only be considered as an order of magnitude estimate.
\label{fig:Mk-m4}}
\end{center}
\end{figure}

In \Figref{fig:Mk-m4} we show the iso-contours of different contributions 
to light neutrino masses in the $m_4 - M$ plane for fixed 
$m=31\,\MeV$ and $m_E=m_4+200\,\GeV$. As in the \Figref{fig:con2}, the contours of the tree level contribution can be obtained from \Eqref{eq:treem} and the contours of the loop contribution can be read off from \Eqref{eq:loopm}. The usual see-saw contribution, $m^\mathrm{ss}=-m^2/M$, does not depend on $m_4$ and the corresponding iso-contours are just vertical lines. The iso-contours are coloured in the same way as in \Figref{fig:con2}. 

Since the loop effect alone cannot explain neutrino data,
the other contributions (the see-saw or/and tree level) should be present 
and without strong suppression. 

For large RH neutrino masses the loop contribution to the light neutrino masses 
dominates over the usual see-saw contribution 
as well as the tree-level contribution. 
In particular, the two loop contribution is
incompatible with RH neutrino masses close to the GUT scale, 
unless the fourth generation effectively decouples, {\it i.e.}~$\xi \lesssim 10^{-12}$. Hence, the 
combination of LH and RH mixing angles of the fourth generation with the three light SM
generations is highly constrained by this contribution. The tree-level
contribution of the fourth generation is only relevant for small RH
neutrino masses (particularly below $100\,\TeV$ for the values
fixed in \Figref{fig:con2}) and therefore small Dirac masses.

Note that contributions of the fourth generation to the light neutrino 
masses (both tree level and loops if $M_4 = M_4^\mathrm{loop}$, see \Eqref{eq:tree} and \Eqref{eq:boundTauTau})  are proportional to 
mixing of the RH neutrinos. Therefore in the limit $U_R \rightarrow 0$, the light neutrino masses are  
generated by the usual see-saw mechanism and the left mixing 
of the fourth generation with the first three generations can be large: at the level of the upper bounds. 

\subsection{3 + 1 Generation Case: an Example}
\label{sec:example}

The results  ``1+1'' generations 
presented in the previous section can be also used in the analysis of 
the (3 + 1) case. 
Here 
we present an example, where the two loop contribution of the
fourth generation is essential for neutrino masses. 

We assume that three massive RH neutrinos have the common Majorana mass
$M_0=10^9 \GeV$ and the Majorana mass of the fourth RH neutrino is zero at tree-level. In
the flavour basis, the neutrino Dirac mass matrix of the three light SM generations is given
by the democratic mass matrix with the common mass scale of
$m=31\,\MeV$, so that the  usual see-saw contribution equals
\begin{equation}
m^\mathrm{ss}=
-0.00288\,
\left(\begin{array}{ccc}
1&1&1\\
1 &1 &1\\
1&1&1\\
\end{array}
\right)\,\eV\;.
\end{equation}
We take the  Dirac mass of the fourth generation to be $m_4=400\,\GeV$ and the
fourth generation charged lepton mass, $m_E=600\,\GeV$. For the RH mixing of
the fourth generation  $(U_R)_{\alpha4}\simeq 1.08\cdot
10^{-4}$ and the left-handed mixing $(U_L) \simeq 1.08\cdot 10^{-4}\cdot
\left(0.15,\,1,\,-1\right)$, the two-loop contribution with a cutoff
scale $\Lambda=10\, M_0$ equals
\begin{equation}
m^\mathrm{loop}=
-0.02473\,
\left(\begin{array}{ccc}
0.0225&0.15&-0.15\\
0.15 &1 &-1\\
-0.15&-1&1\\
\end{array}
\right)\,\eV\;.
\end{equation}
The tree-level contribution of the fourth generation is of the order
of $10^{-6}\eV$, and therefore, negligible.

The loop and see-saw contributions have
 rank 1 and their combination leads to a strong normal
 mass hierarchy with the mass splittings
$\Delta m_{31}^2 = 2.50\cdot 10^{-3}$ eV$^2$ and 
$\Delta m_{21}^2  = 7.41\cdot 10^{-5}$ eV$^2$, 
and mixing angles $\sin^2\theta_{12}=0.330$,  $\sin^2\theta_{13}=0.013$,
and $\sin^2\theta_{23}=0.510$ in agreement with observations.

\section{The Fourth Generation and Symmetries \label{sec:symmetries}} 

In spite of many  efforts to explain the observed features of lepton mixing using 
various discrete flavour symmetries,  no convincing model has been proposed so far 
(see~\cite{Mohapatra:2006gs} for recent review). 
In this connection, we will explore whether the existence of 
4th generation can  help in the realisation of discrete flavour symmetries.  
Existence of four generations of fermions  
can be explained if the flavour symmetry group has the lowest 
irreducible representation \Rep{4} (apart from singlet representations).
The key feature here is very small mass of one right handed neutrino,  
$M_4  \ll M_k$, which can be a consequence of a certain symmetry.  
In this connection, a natural question is whether   
the same symmetry which leads to  $M_4  \ll M_k$
can produce certain  flavour structures for the three light generations? 

In the following, we will present some general results on model-building  
in the four generation context and then focus on the simplest  
symmetry group in detail.

\subsection{General Comments}

As we have shown in \Secref{sec:discussion}, the following features 
are important  for model building:  

\begin{itemize} 

\item The RH neutrino mass matrix should be nearly singular. 
Three massive and one (almost) massless RH neutrino are required.

\item The two loop contribution 
to the light neutrino mass matrix has rank 1 and its 
flavour structure is given by $f_L$, that is,  by the LH mixing of 
the 4th generation with the three light SM generations. 

\item The tree-level contribution to $m_\nu$ produced by mixing of light neutrinos with 
fourth generation neutrino  is negligible compared to the two 
 loop contribution for large RH neutrino masses, and it becomes  
 important only for small $M_k$, as it can be seen in
 \Figref{fig:con2} and \Figref{fig:Mk-m4}. 

\end{itemize} 

The simplest possibility to obtain a pseudo-Dirac structure for the fourth neutrino is to 
impose the conservation of the fourth generation lepton number,  
$L_4$. This implies decoupling of the 4th generation.  
Breaking of the lepton number symmetry is then needed to mix the 4th 
neutrino with the light neutrinos.  
This leads to the tree level contribution $m^{tree}$, and possibly to the generation of a 4th generation Majorana mass term, which in turn produces  a two loop contribution.
The spontaneous breaking of the global $\U{1}_{L_4}$ symmetry results in a Goldstone boson 
(Majoron). This Majoron is not dangerous, because it couples directly 
to the fourth generation only and its coupling with 
the three light SM generations is suppressed:  
$$
g_{\alpha\beta} \sim U_{\alpha 4} U_{\beta 4} M_4/m_4 .  
$$ 
Notice that this  coupling  has a similar dependence  
on $U_{\alpha4}U_{\beta 4} M_4$ as the 2-loop contribution.  
For $M_4 =1$ GeV and $(U_{L})_{\alpha4}^2\sim 10^{-7}$, 
one  obtains $g_{\alpha\beta} \lesssim 10^{-8}$ which satisfies limits on the Majoron 
couplings~\cite{Lessa:2007up}.  
Different values of $M_4$ lead to a similar limit on the coupling $g_{\alpha\beta}$ due to the similar dependence on $U_{\alpha4}U_{\beta 4} M_4$.
The $\U{1}_{L_4}$ symmetry cannot be gauged, 
unless additional particles are introduced to cancel the anomalies.

An abelian symmetry can only forbid certain terms in the 
mass matrix and produce a mass hierarchy  
but cannot lead to relations between different elements of the matrix.  
In this connection, we consider non-abelian groups  
with the lowest non-trivial irreducible representation \Rep{4}.  
This  (i) explains existence of four generations, 
and (ii) opens up a possibility to obtain certain flavour structures.  

We use the SmallGroups catalogue of \texttt{GAP}~\cite{GAP4} 
to obtain the groups  with irreducible representation \Rep{4} 
 in a systematic way. They are   
denoted as    $\SG{N}{m}$, where $N$ is 
the order and $m$ is the index of the group in the SmallGroups catalogue. 
We find  that the smallest group with a 4 
dimensional representation is  
$\SG{20}{3}\cong\mathbb{Z}_5 \rtimes_\varphi \mathbb{Z}_4$. 
It does not contain other non-singlet representations besides \Rep{4} and has order $20$. Hence, it is much smaller than $A_5$, which has been 
studied in~\cite{Chen:2010ty}. 
The next groups with  a real four-dimensional representation are of order $32$: 
\begin{align} 
\SG{32}{6}&\cong((\mathbb{Z}_4 \times \mathbb{Z}_2)\rtimes_\varphi 
\mathbb{Z}_2)\rtimes_\varphi \mathbb{Z}_2 & 
\SG{32}{44}&\cong (\mathbb{Z}_2\times Q_8)\rtimes_\varphi \mathbb{Z}_2\nonumber\\ 
\SG{32}{7}&\cong (\mathbb{Z}_8\rtimes_\psi \mathbb{Z}_2)\rtimes_\varphi\mathbb{Z}_2& 
\SG{32}{49}&\cong(\mathbb{Z}_2\times D_4)\rtimes_\varphi \mathbb{Z}_2\nonumber\\ 
\SG{32}{8}&\cong (\mathbb{Z}_2\times\mathbb{Z}_2)\rtimes_\varphi 
(\mathbb{Z}_4\times \mathbb{Z}_2) & 
\SG{32}{50}&\cong (\mathbb{Z}_2\times Q_8)\rtimes_\varphi \mathbb{Z}_2\nonumber\\ 
\SG{32}{43}&\cong(\mathbb{Z}_2\times D_4)\rtimes_\varphi \mathbb{Z}_2\;, 
\end{align} 
where the defining homomorphism $\varphi$/$\psi$ of each semi-direct product is not  specified explicitly. 
The smallest groups with a complex four-dimensional representation are 
of order $60$: 
\begin{align} 
\SG{60}{6}&\cong\mathbb{Z}_3 \times ( \mathbb{Z}_5 \rtimes_\varphi \mathbb{Z}_4)& 
\SG{60}{7}&\cong\mathbb{Z}_{15}\rtimes_\varphi \mathbb{Z}_4& 
\SG{60}{8}&\cong S_3\times D_{5}\;.
\end{align}
In the following, we will concentrate on the smallest group 
$\SG{20}{3}$.

\subsection{\mathversion{bold} The Smallest Group: 
$\SG{20}{3}\cong\mathbb{Z}_5 \rtimes_\varphi \mathbb{Z}_4$} 

The smallest group with a four-dimensional representation,  $\SG{20}{3}$, 
is the Frobenius  group of order $20$, with presentation 
\begin{equation} 
\left\langle s,t| s^4 = t^5 = \mathbb{1}, ts = st^2\right\rangle,  
\end{equation} 
which can be considered as a subgroup of $S_5$ generated by  
\begin{equation} 
\left\langle (2, 3, 5, 4) , (1, 2, 3, 4, 5)\right\rangle\;.
\end{equation}
The decomposition of the Kronecker product 
$\MoreRep{1}{i}\times\MoreRep{1}{j}$ equals 
\begin{equation}
\left(\begin{array}{cccc}
\MoreRep{1}{1} & \MoreRep{1}{2} & \MoreRep{1}{3} & \MoreRep{1}{4} 
\end{array}\right) \times \left(\begin{array}{c} \MoreRep{1}{1} \\ 
\MoreRep{1}{2} \\ 
\MoreRep{1}{3} \\ 
\MoreRep{1}{4} 
\end{array}\right) 
= \left(\begin{array}{cccc}
\MoreRep{1}{1} & \MoreRep{1}{2} & \MoreRep{1}{3} & \MoreRep{1}{4} \\ 
\dots & \MoreRep{1}{1} & \MoreRep{1}{4} & \MoreRep{1}{3} \\ 
\dots & \dots & \MoreRep{1}{2} & \MoreRep{1}{1} \\ 
\dots & \dots & \dots & \MoreRep{1}{2} \end{array} \right)\;,
\end{equation}
the Kronecker product of $\Rep{4}$ with any of the singlet representations is given by
\begin{equation}
\Rep{4}\times\MoreRep{1}{i}= \Rep{4}\;,
\end{equation}
and the non-trivial Kronecker product of  $\Rep{4}\times \Rep{4}$  is  
\begin{align} \label{eq:KronProd}
\{\Rep{4}\times\Rep{4}\} & = 
\MoreRep{1}{1}\oplus\MoreRep{1}{2}\oplus\Rep{4}_S\oplus\Rep{4}_S & 
[\Rep{4}\times\Rep{4}] &= \MoreRep{1}{3} \oplus\MoreRep{1}{4}\oplus \Rep{4}_A , 
\end{align} where $\{\,\}$ denotes symmetrisation and  $[\,]$ -- antisymmetrisation.  
The other group theoretical details of $\SG{20}{3}$ are summarised 
in \Appref{app:SG20.3}. 

Note that $\SG{20}{3}$ does not contain any 
subgroup with an irreducible representation $\Rep{3}$ because  
20 is not divisible by 3.  
Therefore it cannot be broken down to $\Rep{3} + \Rep{1}$,  
and consequently specific properties of the 4th generation 
compared to the three other generations cannot be explained 
as immediate consequence of the symmetry  breaking.

\subsection{\mathversion{bold}Flavour Structures and $\SG{20}{3}$}

Let us find  possible flavour structures (structures of the fermion mass matrices)  
which can be obtained with $\SG{20}{3}$ symmetry.  
The required  Clebsch-Gordan coefficients are given in \Appref{app:SG20.3}.   

1)  In the limit of exact symmetry, the operators which lead to fermion masses  
have the form $m F_1 F_2$. 
(We omit the usual non-flavoured Higgs fields,  which should be added to satisfy gauge invariance.) 
Here $F_1$ and  $F_2$ are fermion multiplets 
transforming under a certain representation of $\SG{20}{3}$. 
If $F_i$ form quartets, $F_i \sim \Rep{4}$,  
the mass operators has group structure $\Rep{4}\times\Rep{4}$ 
and leads to the symmetric non-singular mass matrix:  
\begin{equation}
\left(\begin{array}{cccc} 0 & 0 & m & 0 \\
 0 & 0 & 0 & m \\
 m & 0 & 0 & 0 \\ 
0 & m & 0 & 0
\end{array}\right). 
\end{equation}
Apparently it cannot be used for RH neutrinos. 
If $F_i$ are singlets of $\SG{20}{3}$,  
$F_i  \sim 
\MoreRep{1}{1}\oplus \MoreRep{1}{2}\oplus 
\MoreRep{1}{3}\oplus \MoreRep{1}{4}$
the mass operators,  $(\MoreRep{1}{1}\oplus \MoreRep{1}{2}\oplus 
\MoreRep{1}{3}\oplus 
\MoreRep{1}{4}) \times(\MoreRep{1}{1}\oplus \MoreRep{1}{2}\oplus 
\MoreRep{1}{3}\oplus \MoreRep{1}{4})$,  
generate the mass matrix 
\begin{equation}
\label{eq:mmatr11}\left(
\begin{array}{cccc} 
m_1 & 0 & 0 & 0 \\
 0 & m_2 & 0 & 0 \\
 0 & 0 & 0 & m_4 \\
 0 & 0 & m_3 & 0\end{array}
\right).
\end{equation}
If one of the mass parameters vanishes, $m_i = 0$,  this matrix has a vanishing eigenvalue.   
It  can be used to describe three massive and one massless 
Majorana RH neutrinos. In the case  $F_1 = F_2 = N$,   
the matrix (\ref{eq:mmatr11})  is symmetric 
and the condition   $m_1 = 0$ or $m_2 = 0$ should be satisfied. 

2) Let us consider operators of the type $y F_1 F_2 \chi$     
with flavon fields  $\chi$  which transform non-trivially  under $\SG{20}{3}$.  
They  generate the mass terms, when the flavour symmetry 
is broken: $\ev{\chi}  \neq 0$.  
We introduce the  
quartet flavons, $\phi \sim \Rep{4}$, 
and the singlets $\chi_i \sim \MoreRep{1}{i}$, with $i=1,2,3,4$, and 
denote the VEVs of these fields as  $\langle \chi_i\rangle  =  
\ev{\MoreRep{1}{i}} = u_i$ and 
$\langle \phi \rangle =  \ev{\Rep{4}}= 
\begin{pmatrix}v_1, &v_2, &v_3, &v_4\end{pmatrix}$. 
If $F_1,  F_2 \sim \Rep{4}$, then the following invariant operators 
can be introduced: 
\be
2\,y_j ( F_1 F_2)_j \chi_j ~~~ j =  2, 3, 4,~~~
y_{S1} \{F_1 F_2\} \phi,~~~\sqrt{2} y_{S2} \{F_1 F_2\} \phi,~~~\sqrt{2} y_{A} [F_1 F_2] \phi,
\label{eq:operator1}
\ee 
where $y_i$ are the Yukawa couplings, and in the second  operator there are three different possibilities  of pairing.   (The operator with $j = 1$ 
gives the structure (\ref{eq:mmatr11}) without symmetry breaking.)  
The  operators  (\ref{eq:operator1}) produce the matrix 
\begin{equation}
\label{eq:matrFF} 
\left(
\begin{array}{cccc}
 v_2 y_{S1} & v_1 (y_{S2}+y_A) & u_2 y_1+u_3 y_3+u_4 y_4 & v_4 (y_{S2}-y_A) \\
 v_1 (y_{S2}-y_A) & v_3 y_{S1} & v_2  (y_{S2}+y_A) & -u_2 y_2-\I u_3 y_3+\I u_4 y_4 \\ 
u_2 y_2 - u_3 y_3 -u_4 y_4 & v_2 (y_{S2}-y_A) & v_4 y_{S1} & v_3  (y_{S2}+y_A) \\
 v_4  (y_{S2}+y_A) & \I \left(u_3 y_3-u_4 y_4\right)-u_2 y_2 & v_3 (y_{S2}-y_A) & v_1 y_{S1}
\end{array}
\right)
\end{equation} 

If $F_1, F_2 \sim \oplus_{i=1}^4\MoreRep{1}{i}$ are 
4 different singlets of the symmetry group,  
then the symmetry structure of the fermionic part of the operator is 
$\oplus_{i=1}^4\MoreRep{1}{i}\times\oplus_{i=1}^4\MoreRep{1}{i}$ 
and only singlet flavon fields can be used. 
The invariant combinations 
\begin{align} y_1 (F_1)_2 (F_2)_1 \chi_2,\quad 
y_2 (F_1)_1 (F_2)_2 \chi_2,\quad 
y_3 (F_1)_3 (F_2)_3 \chi_2,\quad 
y_4 (F_1)_4 (F_2)_4 \chi_2,  \nonumber\\ 
y_5 (F_1)_4 (F_2)_1 \chi_3,\quad 
y_6 (F_1)_3 (F_2)_2 \chi_3,\quad 
y_7 (F_1)_2 (F_2)_3 \chi_3,\quad 
y_8 (F_1)_1 (F_2)_4 \chi_3, 
\\
\nonumber 
y_9 (F_1)_3 (F_2)_1 \chi_4,\quad 
y_{10} (F_1)_4 (F_2)_2 \chi_4,\quad 
y_{11} (F_1)_1 (F_2)_3 \chi_4,\quad 
y_{12} (F_1)_2 (F_2)_4\chi_4 
\end{align} 
generate the mass matrix
\begin{equation}
\left( 
\begin{array}{cccc} 
 0 & u_2 y_2 & u_4 y_{11} & u_3 y_8 \\ 
 u_2 y_1 & 0 & u_3 y_7 & u_4 y_{12} \\ 
 u_4 y_9 & u_3 y_6 & u_2 y_3 & 0 \\ 
 u_3 y_5 & u_4 y_{10} & 0 & u_2 y_4 
\end{array} 
\right)\;.  
\end{equation} 
The value of each matrix  element is independent;  
and if only one flavon field $\chi_i$ is introduced (i.e only one $u_i$ in the matrix above is non-zero) the symmetry only demands that 
four matrix elements are generated. Apparently there is no contribution  
from the flavon $\phi$. 
Finally, if $F_1\sim\oplus_{i=1}^4\MoreRep{1}{i}$ and 
$F_2\sim\Rep{4}$, the fermionic flavour structure $\Rep{4}\times 
\oplus_{i=1}^4\MoreRep{1}{i}$ requires the quartet of flavons 
$\phi$. The operators  
\begin{equation} 
y_1 \{F_2\phi\}_1 (F_1)_1,\quad 
y_2 \{F_2\phi\}_2 (F_1)_2,\quad 
y_3 [F_2\phi]_4 (F_1)_3,\quad 
y_4 [F_2\phi]_3 (F_1)_4 
\end{equation}
produce the mass matrix 
\begin{equation} 
\frac12
\left( 
\begin{array}{cccc} 
 v_3 y_1 & v_4 y_1 & v_1 y_1 & v_2 y_1 \\ 
 v_3 y_2 & -v_4 y_2 & v_1 y_2 & -v_2 y_2 \\ 
 v_3 y_3 & -i v_4 y_3 & -v_1 y_3 & i v_2 y_3 \\ 
 v_3 y_4 & i v_4 y_4 & -v_1 y_4 & -i v_2 y_4 
\end{array} 
\right)\; . 
\end{equation} 

Several important conclusions can be drawn from the forms of these mass matrices. 
A singular Majorana mass matrix with only one vanishing mass eigenvalue  
cannot be obtained as an immediate  result of the $\SG{20}{3}$ symmetry 
or its breaking without additional assumptions or symmetries.  
Such a  matrix can be obtained by  tuning of couplings which in turn 
requires introduction of  additional  symmetries. 
For example,  suppose the RH neutrinos 
transform as $\Rep{4}$ and the direct mass terms are somehow 
forbidden so that  the leading order contribution  comes from the 
one flavon insertion   $\phi \sim  \Rep{4}$.  Then according to \Eqref{eq:matrFF} , one massless RH neutrino  
can be obtained  with the VEV alignment  
$\ev{\phi}= v \left(1,\,1,\,1,\,1\right)$,  and equality  $y_{S1}=\pm y_{S2}$ of 
the Yukawa couplings  of the two possible (symmetric  $\{N N\}$)  invariants. 
A study of the simplest potential with one 
four-dimensional representation shows that the only allowed VEV 
configuration is indeed  $\ev{\phi}=v\left(1,\,1,\,1,\,1\right)$, 
which breaks $\SG{20}{3}$ to $\mathbb{Z}_4$, as it is shown in 
\Tabref{tab:SG20.3},  unless there are special relations between 
parameters in the flavon potential. 

\subsection{Models and Phenomenology} 

In the following, we discuss  the leading order predictions for 
different symmetry assignments for the fields. 
Let us assign for leptons the following transformation properties: 
$\ell\sim\Rep{4}$ and $e_{R}\sim\MoreRep{1}{2}+\MoreRep{1}{3}+\MoreRep{1}{4}+\MoreRep{1}{1}$, 
which allows to generate different charged lepton masses and explain  the number 
of generation.   We use flavons  
$\phi\sim\Rep{4}$ and $\chi_i\sim\MoreRep{1}{i}$. 

1) If the RH neutrinos transform as singlets $N\sim\MoreRep{1}{2}+\MoreRep{1}{3}+\MoreRep{1}{4}+\MoreRep{1}{1}$, the RH Majorana mass matrix has full rank. 
In order to obtain a singular RH neutrino mass matrix, 
we set $M_4$ to zero (which according to our symmetry assignment 
corresponds to parameter $m_1$ in matrix (\ref{eq:mmatr11})). 
This can be obtained in different ways: (i) by choice, (ii) by a ``missing'' representation, {\it i.e.}~by choosing the assignment of representations of $N$ in such a way that there is only one of the complex conjugate representations $\MoreRep{1}{3,4}$ or (iii) by an additional 
symmetry, {\it e.g.}~$N_4\rightarrow \I\, N_4$, which we are going to discuss in the following. This effectively 
leads to a $3+1$ structure of the RH neutrinos and forbids several couplings in the Dirac mass matrices. Therefore, we demand the following transformation properties
$\phi \rightarrow -\I \phi$, $E_4\rightarrow \I E_4$ as well as introduce another flavon $\eta\sim\MoreRep{1}{1}$ transforming as $\eta\rightarrow \I \eta$. All other fields are invariant under the additional symmetry.
The leading order Lagrangian is given by 
\begin{multline} 
-\mathcal{L}= M_1 N_1^T N_1 + 
M_2 \left(N_2^T N_3 + N_3^T N_2\right) + 
Y_{4}^{\nu} \bar \ell N_4 H_1 \frac{\phi}{\Lambda} 
+ Y_{4}^{l} 
\bar \ell H_2 e_{4R} \frac{\phi}{\Lambda}\\
+ 
Y_{k}^{\nu} \bar \ell N_k H_1 \frac{\phi\eta}{\Lambda^2} 
+ Y_{k}^{l} 
\bar \ell H_2 e_{kR} \frac{\phi\eta}{\Lambda^2}  +\hc
\end{multline}
with $k=1,2,3$.
Here $Y_{j}^{l}$  (j = 1, 2, 3, 4) are the charged lepton Yukawa couplings and
$Y_{i}^{\nu}$  (i = 1, 2, 3, 4)  are the neutrino Yukawa couplings.   
This leads to the following $4\times 4$ 
mass matrices in the basis 
$\nu_{e,\mu,\tau,E}$ and  $N_{1,2,3,4}$ for $\ev{\phi}= v 
\left(1,\,1,\,1,\,1\right)$ and $\ev{\eta}=u$: 
\begin{align}
M&= \left(
\begin{array}{cccc}
 M_1 & 0 & 0 & 0 \\
 0 & 0 & M_2 & 0 \\
 0 & M_2 & 0 & 0 \\
 0 & 0 & 0 & 0
\end{array}
\right),& m&=\frac{v\,\vEWnu}{2\Lambda}\left(
\begin{array}{cccc}
 Y_1 \frac{u}{\Lambda} & Y_2 \frac{u}{\Lambda}& Y_3 \frac{u}{\Lambda}& Y_4 \\
 -Y_1\frac{u}{\Lambda} & -i Y_2 \frac{u}{\Lambda}& i Y_3 \frac{u}{\Lambda}& Y_4 \\
 Y_1\frac{u}{\Lambda} & -Y_2 \frac{u}{\Lambda}& -Y_3 \frac{u}{\Lambda}& Y_4 \\
 -Y_1\frac{u}{\Lambda} & i Y_2 \frac{u}{\Lambda}& -i Y_3 \frac{u}{\Lambda}& Y_4
\end{array}
\right)\;. 
\end{align}
The charged lepton mass matrix has the same structure 
as the Dirac neutrino mass matrix. 

In the basis,   where the mass matrices of charged leptons and 
  the RH neutrinos are diagonal:   
$m_e^{fl}=\frac{v\,\vEWe}{2\Lambda}\, 
\diag(Y_{1}^lu/\Lambda,\,Y_{2}^lu/\Lambda,\,Y_{3}^lu/\Lambda,\,Y_{4}^l)$, 
$M^{fl}=\diag(M_1,\,M_2,\,M_2,\,0)$,   
the Dirac neutrino mass matrix becomes
\begin{equation} 
m^{fl}=\frac{v\,\vEWnu}{\Lambda} 
\left(
\begin{array}{cccc}
 Y_1^\nu \frac{u}{\Lambda}& 0 & 0 & 0 \\
 0 & -\frac{i Y_2^\nu u}{\sqrt{2}\Lambda} & \frac{Y_2^\nu u}{\sqrt{2}\Lambda} & 0 \\
 0 & \frac{i Y_3^\nu u}{\sqrt{2}\Lambda} & \frac{Y_3^\nu u}{\sqrt{2}\Lambda} & 0 \\
 0 & 0 & 0 & Y_4^\nu
\end{array}
\right)\;,
\end{equation} 
where $\vEWnu \equiv \ev{H_1}$ and $\vEWe \equiv \ev{H_2}$.
It is not diagonal like the charged lepton mass matrix due 
to the additional rotation from diagonalising the RH neutrino mass matrix.
Its structure corresponds to $f_L = f_R = 0$ in our general consideration of \Secref{sec:NuMass}. 
Hence, the fourth generation decouples from the first three 
generations and the only contribution to the light neutrino mass matrix originates from the ordinary see-saw mechanism
\begin{equation} 
m^\mathrm{ss}=\frac{{\vEWnu}^2 v}{\Lambda^2}\left( 
\begin{array}{ccc} 
 A & 0 & 0 \\ 
 \dots & 0 & B \\ 
 \dots & \dots & 0 \\ 
\end{array} 
\right)\;.
\label{eq:NuMassStructure}
\end{equation}
The coefficients $A$ and $B$ are given by $A=-Y_1^2 \frac{u^2 v}{M_1\Lambda^2} $ 
and $B=-Y_2Y_3\frac{u^2v}{M_3\Lambda^2}$, respectively, and lead to a vanishing 
atmospheric mass squared difference.  

This mass squared difference can be obtained  by the introduction of flavon  $\chi_2\sim\MoreRep{1}{2}$, whose different couplings 
$N_2^T N_2 \chi_2$ and $N_3^T N_3 \chi_2$ split masses of the RH neutrinos $N_2$ and $N_3$ 
which otherwise equal $M_2$ (see $M^{fl}$ above). This split in turn generates 
non-zero elements $m^\mathrm{ss}_{\mu\mu}$ and $m^\mathrm{ss}_{\tau\tau}$  in the see-saw matrix (\ref{eq:NuMassStructure}).
If flavon singlets $\chi_k\sim \MoreRep{1}{k}$ 
with $k=2,3,4$ and $\ev{\chi_k}= u_k$ are introduced, which transform under additional symmetry as $\chi_k\rightarrow-\I\chi_k$, mixing between the fourth generation and the three light generations is generated in the RH Majorana mass matrix by 
interactions $N_4^T (h_2 \chi_2 N_1+ h_4 \chi_4 N_2 + h_3\chi_3 N_3)$ with Yukawa couplings $h_k$. Also these additional flavons contribute to the $3\times 3$ block of the first three generations in the Dirac mass matrices because  $\phi \chi_k^\dagger$ is invariant under the additional symmetry. This leads to appearance of $f_{L,R}$ as well as $M_4$ due to the mixing of $N_4$  with $N_k$. Hence, there are the rank 1 two loop contribution and the  tree level contribution to neutrino masses (see  in \Secref{sec:NuMass}). 
As the tree level contribution is generated at a higher order in flavon insertions compared to the see-saw contribution, it can be neglected at leading order.

Let us comment on other possible VEV alignments 
and structure of the neutrino Dirac mass matrix. 
Any VEV alignment of the 
quartet, which differs from equality of components,  
induces  mixing between the fourth neutrino 
and the three light ones, which is proportional to the deviation 
from the VEV alignment $\ev{\phi}=v\left(1,\,1,\,1,\,1\right)$ 
besides generating the elements $m^\mathrm{ss}_{\mu\mu}$ and $m^\mathrm{ss}_{\tau\tau}$. However, the constraints on the mixing between the fourth and the three light SM generations does not allow large enough values for $m^\mathrm{ss}_{\mu\mu}$ and $m^\mathrm{ss}_{\tau\tau}$ without introducing additional flavons $\chi_k\sim\MoreRep{1}{k}$.

Summarising, the simplest construction with only one flavon $\phi\sim\Rep{4}$ does 
give correct values of neutrino masses and mixing.

2) On the contrary, suppose the RH neutrinos transform as $N\sim\Rep{4}$ and the 
direct mass term $NN$ is forbidden by an additional auxiliary symmetry, {\it e.g.}~$N\rightarrow \omega N$ with $\omega=e^{2\pi \I/3}$ in order to achieve a singular RH neutrino mass matrix, 
the Lagrangian is given by 
\begin{multline} 
-\mathcal{L}= 
\frac12 h_1 \{N^T N\}_\mathrm{diag} \phi  +h_2 \{N^T N\}_\mathrm{off-diag} \phi 
+ Y_{k}^\nu (\bar \ell N )_k H_1 \frac{\phi^\dagger}{\Lambda} 
+ Y_{i}^{l} \bar \ell H_2 e_{i R} \frac{\phi^\dagger}{\Lambda} +\hc\;,
\end{multline} 
where $\phi\sim\Rep{4}$ and $\phi\rightarrow\omega\phi$ as well as $e_{iR}\rightarrow \omega e_{iR}$.
$Y_{k}^\nu$ $(k = S1, S2,  A)$ correspond to three possible combinations 
of  $(\bar \ell N )$ which transform as $\Rep{4}_S$, $\Rep{4}_S$, $\Rep{4}_A$
(see \Eqref{eq:KronProd}). 
Taking the VEV alignment $\ev{\phi}= v \left(1,\,1,\,1,\,1\right)$, 
the RH neutrino mass matrix becomes 
\begin{equation} 
M= \frac{v}{4}\left( 
\begin{array}{cccc}   
h_1 & h_2 & 0 & h_2 \\  
\dots & h_1 & h_2 & 0 \\   
\dots & \dots & h_1 & h_2 \\
 \dots & \dots & \dots & h_1 
\end{array}
\right)  
\end{equation}  
with mass eigenvalues $|h_1| v$, $|h_1|v$, $|h_1-2h_2|v$ 
and $|h_1+2h_2|v$. Hence, there is exactly one massless RH neutrino 
if $h_1=\pm2h_2$. Under the assumption $h_1= -2h_2$, 
which can be obtained  by fine-tuning the couplings 
$h_{1,2}$ only, the diagonalised RH neutrino mass matrix becomes  
$M^{fl}=|h_1| v\,\diag(2,1,1,0)$.  The eigenstate corresponding
to the zero  mass eigenvalue is $\left(1,\,1,\,1,\,1\right)$. 
In flavour basis, where the charged leptons and the RH neutrino mass matrix are diagonal, the charged lepton mass matrix is  
$m_e^{fl}=\frac{v\,\vEWe}{2\Lambda}\, 
\diag(Y_1^e, \, Y_2^e,  \, Y_3^e,  \,Y_4^e)$ and the Dirac neutrino mass matrix is given by
\begin{equation}
m^{fl} =\frac{v\,\vEWnu}{2\Lambda}\left(
\begin{array}{cccc}
 2 Y^\nu_{S1}-Y^\nu_{S2} & 0 & 0 & 0 \\
 0 & -\frac{2 Y^\nu_A+i Y^\nu_{S2}}{\sqrt{2}} & \frac{2 i Y^\nu_A-Y^\nu_{S2}}{\sqrt{2}} & 0 \\
 0 & \frac{-2 Y^\nu_A+i Y^\nu_{S2}}{\sqrt{2}} & \frac{-2 i Y^\nu_A-Y^\nu_{S2}}{\sqrt{2}} & 0 \\
 0 & 0 & 0 & 2 Y^\nu_{S1}+Y^\nu_{S2}
\end{array}
\right)\;.
\end{equation}
This leads  to a see-saw contribution to the neutrino mass matrix of the form (\ref{eq:NuMassStructure}) with coefficients $A=-\frac{(Y_{S2}-2 Y_{S1})^2}{2 h_1}$ and $B=-\frac{4 Y_A^2 + Y_{S2}^2}{h_1}$.  Hence, it has the same structure with different 
coefficients and, essentially, the same conclusions can be drawn as in case 1). 
Similarly to the previous assignment of representations, the addition of flavons $\chi_k\sim\MoreRep{1}{k}$ will lead to a non-vanishing atmospheric mass squared difference and a mixing of the fourth generation with the first three SM generations.

Concluding, the simplest group \SG{20}{3} does not immediately 
lead to the required  flavour structure. For instance  tuning of  Yukawa couplings 
is required to obtain zero mass for one of the RH neutrinos 
if they transform as \Rep{4}. In this case, the direct mass term of a four 
dimensional representation always has full rank, and the RH neutrino 
mass matrix has to be generated using non-singlet flavons. 
Furthermore,  an additional leading order contribution 
to the neutrino mass matrix is required which generates the $(m_\nu)_{\mu\mu}$ and $(m_\nu)_{\tau\tau}$ entries.  We have demonstrated how those contributions can be achieved.

A survey of all small groups  up to order $56$ with the RH neutrinos 
transforming as \Rep{4} shows that the RH neutrino masses are 
either all of the same order of magnitude or there are two heavy and 
two light RH neutrinos. Hence, one might argue that \Rep{4} is generally not the 
best representation for the RH neutrinos and a more viable choice is 
$\Rep{3}\oplus\Rep{1}^\prime$,  where \Rep{3} is a real representation of a given group
and \Rep{1}$^\prime$ a complex representation such that the matrix of direct 
mass terms  is singular with one vanishing mass. This splitting might 
also be obtained from breaking the symmetry group to a smaller subgroup. 
The smallest group, which allows the decomposition  
of \Rep{4} into $\Rep{3}\oplus\Rep{1}$, 
is $A_5$. However $A_5$ has irreducible representations 
\Rep{3}  as well as \Rep{5}, so  that use of representation \Rep{4} only should be justified.  

\section{Summary and Conclusions
\label{sec:conclusions}} 

1. We have explored the generation of light neutrino masses 
in the presence  of a fourth family of fermions with four RH neutrinos 
(1 per family).  In this context, generically there are three contributions 
to the light neutrino masses from
three different mechanisms:  

\begin{itemize}
\item[(i)]
the usual high mass scale see-saw contribution;   
\item[(ii)]
the tree level contribution induced by  mixing 
of the light generations with the fourth generation.  This 
contribution requires mixing of both left and right neutrino   
components ($f_L \neq 0$ $f_R \neq 0$) in the basis 
where the Majorana mass matrix of the RH components is diagonal. 
\item[(iii)] The two loop contribution with two $W$-bosons exchange,  
related to  the non-zero Majorana mass of the fourth neutrino, $M_4$. 
 \end{itemize}
 
2. We show that even if $M_4 = 0$ initially  at tree level, 
it will be generated at the two loop level due to usual Yukawa interactions. 
This radiatively generated mass is proportional to the  
large Majorana masses $M_i$ and therefore, is rather large:  
$0.1 - 1$  GeV. Unless there is strong cancellation,  
({\it e.g.},  with tree level contribution), 
this mass, in turn,   produces the dominant contribution 
to the light  neutrino masses in large part  of parameter space. 
In the case $f_R = 0$, the new contributions related to  the fourth generation 
vanish.  The relative contributions from different mechanisms depend strongly on 
$M, m,  \xi \equiv (U_{L})_{\alpha 4} (U_R)_{i4}$ , and to a smaller degree on $m_E$ and $m_4$. 
The tree level contribution of the fourth generation  dominates over 
the loop contribution for small RH neutrino masses $M$ and vice versa, 
as it is illustrated in Figs. \ref{fig:con2} and \ref{fig:Mk-m4}. 
The smaller Dirac mass, $m$, the larger the tree level contribution compared 
to the usual see-saw contribution; the loop contribution is independent of $m$. 
The usual see-saw contribution does not depend on $\xi$, while 
the tree level contribution and  the loop contribution 
are proportional to $\xi$  and $\xi^2$  respectively.  

3. In general, the contributions from three different mechanisms 
have different flavour structures.   
The loop contribution is singular and therefore it cannot explain the 
observed mass hierarchy. Therefore, comparable contributions should follow 
from other  mechanisms. 
The combination  of the loop and  see-saw contributions  is realized at  
$M_i \sim (10^{7} - 10^{10})$ GeV, a large Dirac neutrino mass $m$,  
and $\xi = 10^{-9} - 10^{-7}$.  
Combination  of  the loop and  tree level  contributions works for  
$M_i \lesssim 10^{5}$ GeV (and therefore small $m$) and $\xi = 10^{-6}$.  
An interplay of  the ``see-saw and tree level'' contributions  
is realized at small Dirac neutrino mass $m$ and $\xi < 10^{-7}$.  
The loop contribution gives very strong bound on mixing parameters  
especially for large $M_i$. At $M_i >  10^{4}$ GeV the bounds are much stronger than those  
from the direct searches. 
  The loop contribution 
can be suppressed  if certain cancellation occurs between  
contributions from different $M_i$ or between the loop  
and tree level contribution, although this looks rather unnatural.   

The upper bound on the light neutrino masses (which follows from cosmology) 
gives the most stringent  bound on the  parameters of the model.

4. In the see-saw limit of the fourth generation, where the  Majorana mass $M_4$
is much larger than the  Dirac mass  
$m_4$,  $M_4 \gg m_4$ , the tree-level contribution 
of the fourth  generation to the three light 
active neutrinos is negligible.    
In the pseudo-Dirac limit,  $m_{4}\gg M_4$, the tree-level contribution 
of the 4th generation to the light neutrino mass 
matrix  is significant for small RH neutrino masses $M_i$ and determines
one of the mass scales.  

5. We explored flavour symmetries, which could explain the 
leptonic flavour structure and studied the smallest group 
with a four-dimensional representation,   
$\SG{20}{3}\cong\mathbb{Z}_5 \rtimes_\varphi \mathbb{Z}_4$. 
We have found the simplest flavour structures (mass matrices), which can be obtained 
as a result of this flavour symmetry. The required singularity 
of the RH neutrino mass matrix  
can be achieved imposing conditions on the Yukawa 
couplings and VEVs.  There is no viable model based on one flavon $\phi\sim\Rep{4}$, but there are phenomenologically viable models can be constructed  with  
flavons $\phi\sim\Rep{4}$ and $\chi_k\sim\MoreRep{1}{k}$.
We indicated the next smallest groups, which might be interesting to study.  

\section*{Acknowledgements}
We thank M.~Lindner for useful discussions in the initial stage of this work.
M.S.~would like to thank J.~Lopez-Pavon for useful discussions
about the $\beta\beta_{0\nu}$-decay, as well as C.~Duhr
and G.~Luisoni for helpful discussions on the evaluation 
of two loop integrals. We would like to thank S. Bhattacharya for useful comments on the first version and, especially, A. Aparici, J. Herrero-Garcia, A. Santamaria and N. Rius for pointing out an error in the two loop calculation. A. Yu. S. acknowledges support by the Alexander von Humboldt Foundation. M.S.~would like to acknowledge MPI f\"ur Kernphysik as well as the ICTP, where a part of
this work has been done, for hospitality of its staff and the generous
support. This work was supported in part by the Australian Research Council.

\appendix
\section{\mathversion{bold}Group Theory of $\SG{20}{3}\cong\mathbb{Z}_5\rtimes_\varphi
  \mathbb{Z}_4$\label{app:SG20.3}}

\begin{table}[htbp]
\begin{center}
\subtable[\,Character table]{\begin{tabular}{l|ccccc}
&\multicolumn{5}{|c}{classes}\\\hline
& $\mathcal{C}_1$& $\mathcal{C}_2$& $\mathcal{C}_3$& $\mathcal{C}_4$& $\mathcal{C}_5$\bigstrut\\\hline
$G$ & $\mathbb{1}$ & $s$& $s^2$& $t$& $s^3$\bigstrut\\\hline
$h_{\mathcal{C}_i}$ &       $1$& $4$& $2$& $5$& $4$\bigstrut\\\hline\hline
$\MoreRep{1}{1}$   &  $1$ & $1$ & $1$ & $1$ & $1$\bigstrut[t]\\
$\MoreRep{1}{2}$   &  $1$ & $-1$ &  $1$ & $1$ & $-1$\\
$\MoreRep{1}{3}$   &  $1$ &  $-\I$ &  $-1$ &   $1$ &  $\I$\\
$\MoreRep{1}{4}$   &  $1$ & $\I$ & $-1$  &  $1$ & $-\I$\\
$\Rep{4}$   &  $4$ & $0$ &  $0$ &  $-1$  & $0$\\
\end{tabular}}
\hspace{4ex}
\subtable[\,Breaking patterns]{\begin{tabular}{l|l}
subgroup & VEV configuration\\\hline\hline
$D_{5}$ & $\ev{\MoreRep{1}{2}}$\bigstrut\\\hline
$\mathbb{Z}_5$ & $\ev{\MoreRep{1}{3}}$\bigstrut[t]\\
$\mathbb{Z}_5$ & $\ev{\MoreRep{1}{4}}$\bigstrut[b]\\\hline
$\mathbb{Z}_4$ &              $\ev{\Rep{4}}\sim\begin{pmatrix} 1,& 1,& 1,& 1 \end{pmatrix}$\bigstrut[t]\\
$\mathbb{Z}_4$ &              $\ev{\Rep{4}}\sim\begin{pmatrix} \eta^4,& \eta,& \eta^2,& 1 \end{pmatrix}$\\
$\mathbb{Z}_4$ &              $\ev{\Rep{4}}\sim\begin{pmatrix} \eta^3,& \eta^2,& \eta^4,& 1 \end{pmatrix}$\\
$\mathbb{Z}_4$ &              $\ev{\Rep{4}}\sim\begin{pmatrix} \eta^2,& \eta^3,& \eta,& 1 \end{pmatrix}$\\
$\mathbb{Z}_4$ &              $\ev{\Rep{4}}\sim\begin{pmatrix}
    \eta,& \eta^4,& \eta^3,& 1 \end{pmatrix}$\bigstrut[b]\\\hline
$\mathbb{Z}_2$ &
$\ev{\Rep{4}}\sim\begin{pmatrix}1,&0,&1,&0\end{pmatrix}$, 
$\begin{pmatrix}0,&1,&0,&1\end{pmatrix}$\bigstrut[t]\\
$\mathbb{Z}_2$ &
           $\ev{\Rep{4}}\sim\begin{pmatrix} \eta^2,& 0,& 1,&
               0\end{pmatrix}$,
$\begin{pmatrix} 0,& \eta,& 0,& 1 \end{pmatrix}$\\
$\mathbb{Z}_2$ & 
$\ev{\Rep{4}}\sim\begin{pmatrix} \eta^4,& 0,& 1,& 0 \end{pmatrix}$, 
$\begin{pmatrix} 0,& \eta^2,& 0,& 1 \end{pmatrix}$\\
$\mathbb{Z}_2$ &
$\ev{\Rep{4}}\sim\begin{pmatrix} \eta,& 0,& 1,& 0 \end{pmatrix}$, 
$\begin{pmatrix} 0,& \eta^3,& 0,& 1 \end{pmatrix}$\\
$\mathbb{Z}_2$ &
$\ev{\Rep{4}}\sim\begin{pmatrix} \eta^3,& 0,& 1,& 0 \end{pmatrix}$, 
$\begin{pmatrix} 0,& \eta^4,& 0,& 1 \end{pmatrix}$\bigstrut[b]\\
\end{tabular}
}

\subtable[\,Generators of \Rep{4}]{
\begin{minipage}{6cm}
\begin{align*}
s&=\left(
\begin{array}{cccc}
 0 & 0 & 0 & 1 \\
 0 & 0 & 1 & 0 \\
 1 & 0 & 0 & 0 \\
 0 & 1 & 0 & 0
\end{array}
\right)&
t&=
\left(
\begin{array}{cccc}
 \eta & 0 & 0 & 0 \\
 0 &  \eta^4 & 0 & 0 \\
 0 & 0 & \eta^2 & 0 \\
 0 & 0 & 0 & \eta^3
\end{array}
\right)
\end{align*}
\end{minipage}
}

\caption{Group theoretical details of $\SG{20}{3}$: Character table in
  Tab.~(a), breaking patterns in Tab.~(b), and the generators of
  \Rep{4} in Tab.~(c). $G$ denotes the
  generating element, $h_{\mathcal{C}_i}$ is the order of the elements 
and $\eta=\exp(2\pi\I/5)$ is the fifth root of unity.  \label{tab:SG20.3}}
\end{center}
\end{table}

The generators of the four-dimensional representation and the character
table are given in \Tabref{tab:SG20.3}. The Clebsch-Gordan coefficients 
for the Kronecker product $4\times4$
can be calculated following the algorithm described
in~\cite{PhysStatSol.b.90.211}. The Kronecker product of 
$a=\left(a_1,\,a_2,\,a_3,\,a_4\right)\sim\Rep{4}$ and
$b=\left(b_1,\,b_2,\,b_3,\,b_4\right)\sim\Rep{4}$ results in
\begin{align}
\MoreRep{1}{1}&\sim \frac12\left(a_ 1 b_ 3+a_3 b_ 1+a_2 b_ 4+a_ 4 b_ 2\right)\,,&
\MoreRep{1}{2}&\sim \frac12\left(a_ 1 b_ 3+a_3 b_ 1-a_2 b_ 4-a_ 4 b_ 2\right)\,, \\\nonumber
\MoreRep{1}{3}&\sim \frac12\left(a_ 1 b_3-a_ 3 b_ 1+\I\, a_ 2 b_ 4-\I\,  a_ 4 b_ 2\right)\,,&
\MoreRep{1}{4}&\sim \frac12\left(a_ 1 b_ 3- a_ 3 b_ 1-\I\,  a_ 2 b_ 4+\I\,  a_ 4 b_ 2\right)\,,
\end{align}
\begin{align}
\Rep{4}_S&\sim \frac{1}{\sqrt{2}}\begin{pmatrix}a_ 3 b_ 4+a_ 4 b_ 3,&a_ 4 b_ 1+
    a_1 b_4,&a_ 1 b_ 2+a_2 b_1,&a_ 2 b_ 3+a_3 b_2\end{pmatrix}\,,\\\nonumber
\Rep{4}_S&\sim \begin{pmatrix}a_ 2 b_ 2,&a_ 3 b_ 3,&a_ 4 b_ 4,&a_ 1 b_ 1\end{pmatrix}\,,\\\nonumber
\Rep{4}_A&\sim \frac{1}{\sqrt{2}}\begin{pmatrix}a_ 3 b_ 4-a_ 4 b_ 3,&a_ 4 b_ 1-
    a_1 b_4,&a_ 1 b_ 2-a_2 b_1,&a_ 2 b_ 3-a_3 b_2\end{pmatrix}\;.
\end{align}
The Clebsch-Gordan
coefficients for $\MoreRep{1}{i}\times \Rep{4}$ with
$\alpha\sim\MoreRep{1}{i}$ and
$\left(b_1,\,b_2,\,b_3,\,b_4\right)\sim\Rep{4}$ are
\begin{align}
\MoreRep{1}{1}\times \Rep{4}&\simeq\begin{pmatrix}\alpha\,  b_1,&\alpha\,  b_2,&\alpha\,  b_3, 
& \alpha\,  b_4\end{pmatrix}\,,&
\MoreRep{1}{2}\times \Rep{4}&\simeq\begin{pmatrix}\alpha\,  
b_1,&-\alpha\,  b_2,&\alpha\,  b_3,&-\alpha\,  b_4\end{pmatrix}\,,\\\nonumber
\MoreRep{1}{3}\times \Rep{4}&\simeq\begin{pmatrix}\alpha\,  
b_1,&-\I\, \alpha\,  b_2,&-\alpha\,  b_3,&\I\, \alpha\,  b_4\end{pmatrix}\,,&
\MoreRep{1}{4}\times \Rep{4}&\simeq\begin{pmatrix}\alpha\,  
b_1,&\I\, \alpha\,  b_2,&-\alpha\,  b_3,&-\I\, \alpha\,  b_4\end{pmatrix}\;,
\end{align}
where $S$ and $A$ indicate that the representation is in the symmetric
or antisymmetric part, respectively.
The different possible breaking patterns of $\SG{20}{3}$ to its
subgroups are shown in \Tabref{tab:SG20.3}, where we have used the algorithm described in
Appendix C of~\cite{Parattu:2010cy}.

\bibliography{4gen}
\end{document}